\documentclass[final,5p,times,twocolumn,sort&compress]{elsarticle}

\usepackage{etoolbox}

\usepackage{amssymb}
\usepackage{amsmath}
\usepackage{empheq}
\usepackage{comment}
\usepackage{balance}
\usepackage{placeins}
\usepackage{enumerate}
\usepackage{subcaption}
\usepackage{graphicx}
\usepackage[inline,shortlabels]{enumitem}
\usepackage[usenames,dvipsnames]{color}
\usepackage{xcolor}
\usepackage[colorlinks=true,linkcolor=Maroon,citecolor=Maroon,urlcolor=Maroon]{hyperref}

\usepackage{xr}         
\makeatletter
\def\ps@pprintTitle{%
 \let\@oddhead\@empty 
 \let\@evenhead\@empty
 \def\@oddfoot{}%
 \let\@evenfoot\@oddfoot}
 \makeatother

\makeatletter
\patchcmd{\@twocolumntrue}{\@twocolumnfalse}{}{}{}
\makeatother

\newcommand{\A}{\mathbf{A}} 			     
\newcommand{\J}{\mathbf{J}}                  
\newcommand{\PP}{\mathbf{P}} 			     
\newcommand{\Q}{\mathbf{Q}} 			     
\newcommand{\U}{\mathbf{U}} 			     
\newcommand{\V}{\mathbf{V}} 			     

\newcommand{\x}{\mathbf{x}} 			     
\newcommand{\z}{\mathbf{z}} 			     
\newcommand{\f}{\mathbf{f}} 			     
\newcommand{\W}{\mathbf{W}} 			     
\newcommand{\abf}{\mathbf{a}} 			     
\newcommand{\bb}{\mathbf{b}} 			     
\newcommand{\vv}{\mathbf{v}} 			     
\newcommand{\p}{\mathbf{p}} 			     
\newcommand{\rr}{\mathbf{\hat{r}}}
\newcommand{\nn}{\mathbf{\hat{n}}}

\newcommand{\et}{\boldsymbol{\eta}}          
\newcommand{\Sig}{\boldsymbol{\Sigma}}       

\newcommand{\SM}{(\textit{SM})}

\begin{document}

\begin{frontmatter}

    \title{Phase Transitions Without Instability:\\ A Universal Mechanism from Non-Normal Dynamics}
    \author[add1]{Virgile Troude}
    \author[add1]{Didier Sornette\corref{cor1}}

    \address[add1]{\scriptsize
        Institute of Risk Analysis, Prediction and Management (Risks-X),\\
        Academy for Advanced Interdisciplinary Sciences,\\
        Southern University of Science and Technology, Shenzhen, China
    }
    \cortext[cor1]{Corresponding author. Email: dsornette@ethz.ch}

    \begin{abstract}
        We identify a new universality class of phase transitions that arises in \emph{non-normal} systems,
        challenging the classical view that transitions require eigenvalue instabilities.
        In traditional bifurcation theory, critical phenomena emerge when spectral stability is lost;
        here, we show that transitions can occur even when all equilibria are spectrally stable.
        The key mechanism is the transient amplification induced by non-orthogonal eigenvectors:
        noise-driven dynamics are enhanced not by lowering energy barriers, but by increasing the effective shear of the flow,
        which renormalizes fluctuations and acts as an emergent temperature. Once the non-normality index $\kappa$ exceeds a critical threshold $\kappa_c$,
        stable equilibria lose practical relevance, enabling escapes and abrupt transitions despite preserved spectral stability.  
        This \emph{pseudo-criticality} generalizes Kramers' escape beyond potential barriers, providing a fundamentally new route to critical phenomena.
        Its implications are broad: in biology, DNA methylation dynamics reconcile long-term epigenetic memory with rapid stochastic switching;
        in climate, ecology, finance, and engineered networks, abrupt tipping points can arise from the same mechanism.  
        By demonstrating that phase transitions can emerge from non-normal amplification rather than eigenvalue instabilities,
        we introduce a predictive, compact framework for sudden transitions in complex systems,
        establishing non-normality as a defining principle of a new universality class of phase transitions.
    \end{abstract}
    
\end{frontmatter}

\section*{Introduction}

    One might expect landscapes to evolve slowly over geological timescales,
    shaped gradually by tectonic deformation and erosion, 
    yet mountains collapse in seconds through catastrophic landslides \cite{hungr2018review}
    or release centuries of accumulated stress in a single earthquake \cite{Burbank11}.
    Social norms are thought to drift imperceptibly through cultural evolution,
    but history shows that revolutions and collective shifts in belief can overturn societies almost overnight.
    Engineered structures such as bridges seem destined to degrade predictably under long-term wear,
    yet their failure can occur abruptly when small defects cascade into instability.
    Financial markets, often assumed to progress steadily alongside economic development,
    can crash within days, as in the 2008 global financial crisis,
    when herding and positive feedback loops transformed confidence into panic \cite{Sornette2017,sornette2023non}.
    Climate is expected to change gradually over millennia,
    yet records reveal the sudden onset of glacial and interglacial periods~\cite{Alley2003}
    or rapid desertification events \cite{Renssen2022}.
    Genetic and epigenetic evolution are presumed to unfold over many generations,
    but organisms can undergo swift regulatory reprogramming,
    including fast DNA methylation transitions under stress~\cite{foster2007stress,busto2020stochastic}.
    Ecosystems appear to adapt smoothly to environmental change,
    yet forests can transform abruptly into savannas \cite{Hirota2011}
    or entire marine food webs can collapse under overfishing \cite{Daskalov2007}.
    Even the solid Earth, seemingly the epitome of slow transformation,
    reveals through volcanic eruptions \cite{Newhall1996}
    and other geophysical outbursts how accumulated energy may discharge in a moment.

    The prevailing explanation for such abrupt transformations invokes the paradigm of tipping points, bifurcations, and critical transitions.
    In this view, systems evolve smoothly until a control parameter crosses a threshold, beyond which the current equilibrium 
    becomes unstable forcing a jump to a new state.
    This framework, which is rooted in the mathematics of bifurcations, has proved remarkably fruitful, linking phenomena as diverse as financial crashes, ecosystem collapses, epileptic seizures, and climate tipping events under a unified picture of instability-driven change.
    Yet this classical view rests on a key assumption: that sudden transitions require the loss of stability of an equilibrium.
    By this logic, a system can only switch states if its equilibrium becomes unstable,
    as if ``instability'' were a prerequisite for change.
    Here, we show that this is not the whole story: phase transitions can occur even when all equilibria remain spectrally stable.
    The only requirements are generic non-variational dynamics and the presence of asymmetric shear,
    which is a condition ubiquitous in natural and engineered systems.
    This observation opens a fundamentally new perspective on critical phenomena, revealing a class of stable phase transitions driven by non-normal amplification rather than eigenvalue instabilities. Across these domains, systems that seem stable and evolving gradually
    reveal a hidden potential for sudden reorganization, which is 
    the empirical signature of what we term \emph{pseudo-critical phase transitions},
    where change arises not from spectral instability but from non-normal amplification within stable dynamics.

  \refstepcounter{figure} 
    \begin{figure*}
    \centering
    \includegraphics[width=\textwidth]{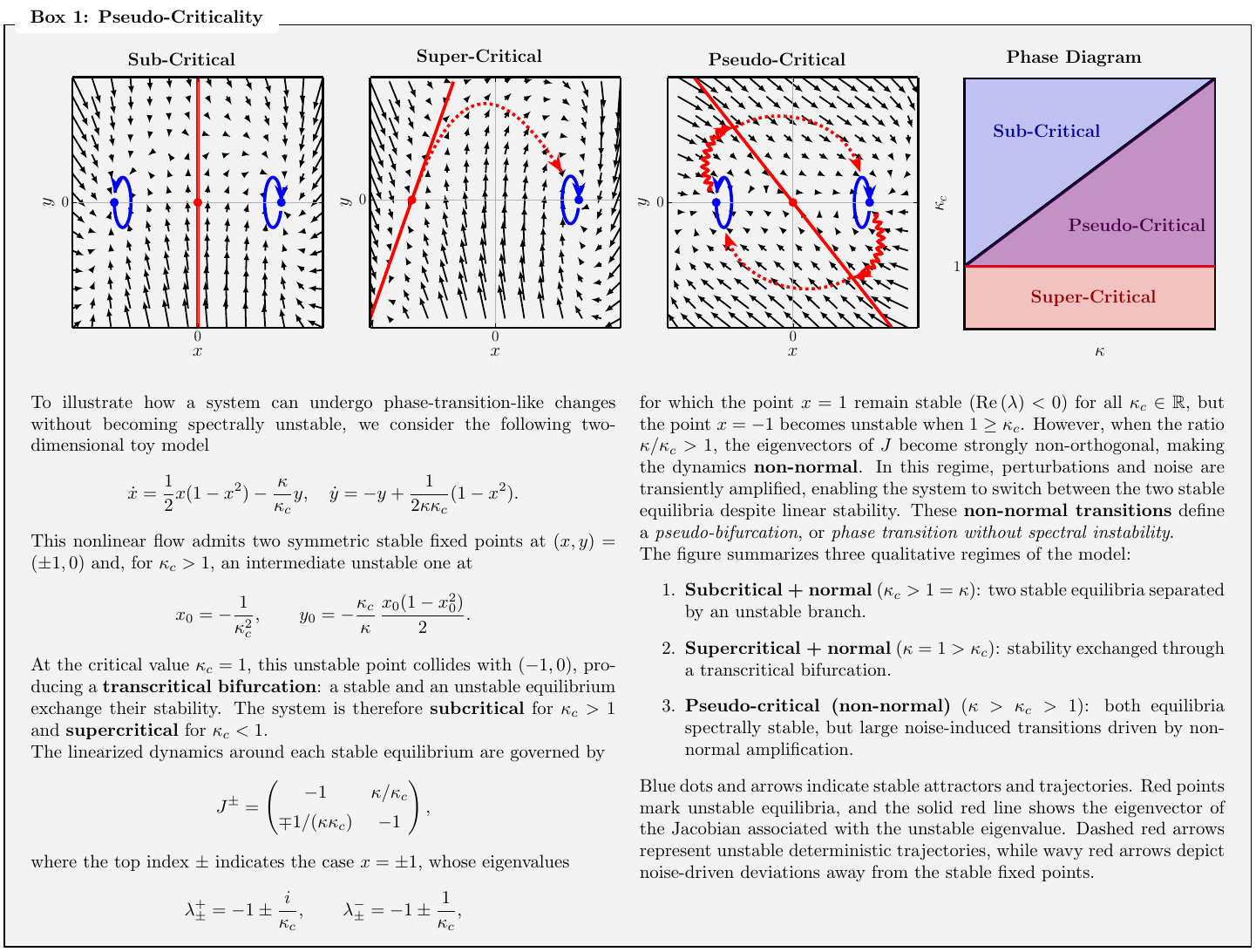}
    \label{box:phase_transition}
    \end{figure*}

    The key novelty is that non-normality -- defined as the existence of non-orthogonal eigenvectors
    of the linearised dynamics -- induces a new form of phase transition. 
    While classical critical phenomena are tied to spectral instabilities (eigenvalues crossing zero real part), 
    we identify a critical threshold of non-normality, $\kappa_c$, 
    beyond which fluctuations are effectively rescaled and escape dynamics undergo a qualitative change. 
    This pseudo-critical transition is not due to eigenvalue instability, 
    but arises from transient amplification of noise: the system escapes from a stable equilibrium 
    because the effective noise level $\delta_{\text{eff}} = \delta(\kappa/\kappa_c)^2$ 
    becomes much larger than the intrinsic noise $\delta$. 
    In this way, non-normality plays the role of an effective temperature, 
    dramatically accelerating transition rates even when spectral stability remains intact.
    To formalize this mechanism, we extend Kramers' theory \cite{kramers1940,melnikov1991kramers}
    into the non-variational regime, obtaining a compact analytical expression for transition probabilities in which all effects of transient amplification are encapsulated by a single parameter, the non-normality degree $\kappa$.
    This generalization unifies previously fragmented insights into fluctuation amplification and yields closed-form escape-rate predictions that can be directly compared with simulations and empirical data.

    The new concept is summarized in Box~1 using a toy model that highlights how our formalism unifies and extends the entire domain of phase transitions within a broader theoretical structure.
    In particular, the phase diagram reveals a wide pseudo-critical region, which is the domain in which our concept of pseudo-critical phase transitions applies.

    Finally, we apply our framework to DNA methylation, a cornerstone of epigenetic regulation. 
    Classical bistable models of CpG dyads~\cite{zagkos2019} capture long-term stability of methylation patterns 
    but fail to explain the observed rapidity of methylation switching~\cite{busto2020stochastic}. 
    We show that, by extending such models to include explicit non-normality, 
    one can reconcile bistability with fast transitions, 
    thus providing a mechanistic explanation for rapid epigenetic responses at physiological temperature. 
    This illustrates the broad relevance of our theory, 
    which not only uncovers a new type of phase transition induced by non-normality, 
    but also provides a unifying principle for accelerated dynamics across physics,
    biology, and the social sciences \cite{sornette2023non}.
    
\section*{From Linear Non-Normal Amplification to Non-Linear Escape Dynamics}

    Before turning to nonlinear escape rates,
    it is useful to situate our contribution within the broader context of non-normal amplification that we,
    and others, have developed in previous work.
      We build on prior extensions of critical phenomena theory to non-normal systems, 
    initiated in the study of hydrodynamic instabilities and atmospheric flows~\cite{Farrell1988,Farrell1989,FarrellIoannou1996a,FarrellIoannou1996b,FarrellIoannou2003,FarrellIoannou2007}. 
    These studies showed that non-normality can generate transient amplification closely resembling behavior near bifurcations. 
    Here, we generalize this perspective well beyond hydrodynamics, 
    establishing a unified nonlinear framework for escape dynamics in arbitrary non-variational, non-normal systems.

    Previous work \cite{troude2024} demonstrated that linear stochastic systems with non-normal Jacobians can exhibit transiently repulsive dynamics
    -- termed pseudo-bifurcations --
    even when all eigenvalues remain stable (negative real parts).
    Such transient behaviour reproduces classic early-warning signals,
    including dimension reduction, increased variance, and critical slowing down,
    without the need for a true bifurcation.

    Building on this, the present authors \cite{troude2025}  introduced a unifying linear framework that combines three canonical mechanisms of amplification
    -- spectral criticality, resonance, and non-normality --
    via two control parameters: (i) the distance to a critical bifurcation or resonance,
    and (ii) the non-normality index (or condition number of the eigenbasis of the Jacobian matrix) \(\kappa\).
    Closed-form expressions were derived quantifying how system response scales with noise or forcing,
    and identified a broad pseudo-critical regime that arises for large \(\kappa\) even far from spectral instabilities.
   This formulation defines a critical degree of non-normality $\kappa_c$:
   the system is pseudo-critical when $\kappa>\kappa_c$,
   while in the approach to genuine spectral criticality one finds $\kappa_c\to 1^+$. 

    While the linear analysis lays the foundation,
    it is insufficient to describe actual escape dynamics in nonlinear multimodal systems,
    such as the rate of switching between stable states.
    To capture these rare noise-driven transitions, we turn to the large-deviation formalism
    (Kramers theory extended to non-normal systems) and derive the effective action and escape rate by
    (i) embedding non-normal coupling into the small-noise reduction,
    (ii) calculating the minimal action path,
    and (iii) obtaining explicit rate formulas in low dimensions.
    Through this derivation, we demonstrate that non-normality renormalizes the noise amplitude,
    effectively raising the ``activation temperature'' and accelerating escape rates,
    even though the deterministic stability landscape remains unchanged.
    This generalization bridges the linear insight of pseudo-critical amplification with rigorous estimates of transition probabilities in fully nonlinear stochastic dynamics.

    We therefore propose the following conceptual progression: the linear regime reveals how non-normality enhances transient responses and amplifies fluctuations \cite{troude2024,troude2025}; the nonlinear large-deviation regime demonstrates how this amplified variability accelerates state switching through a 
    renormalized Kramers escape rate; and our theoretical developments, together with their detailed derivations in the Supplementary Materials, 
    establish the mathematical continuity linking these two regimes.
    We now outline the main steps leading to our key results and claims.    

\section*{Model}

    In numerous (overdamped) physical, social, or biological systems,
    the dynamics of a state $x$ (defined here as a scalar) can be modeled by a Langevin equation
    \begin{equation}
        \dot{x} = f(x) + \sqrt{2\delta} \eta, \quad \eta \overset{iid}{\sim} N(0, 1),
    \end{equation}
    where $f(x)$ represents the force and $\delta$ quantifies the noise amplitude.
    In physical contexts where the noise is due to thermal fluctuations,
    $\delta = k_B T$, where $k_B$ is Boltzmann's constant and $T$ denotes the system's temperature.
    Under the assumption that the force is smooth, we can write it as the derivative of a potential,
    i.e., $f(x) = -\phi'(x)$, which defines the framework of the classical Kramers escape problem \cite{kramers1940,melnikov1991kramers}:
    a representative point at position $x$ moves in one dimension in an (energy) potential $\phi(x)$.
    For a potential with a local minimum at $x_i$ and a barrier (local maximum) at $x_f$,
    the escape rate $\Gamma$ is given by (see \textit{Supplementary Materials} (SM))
    \begin{equation}    \label{eq:kramer_classical}
        \Gamma = \frac{1}{2\pi} \sqrt{\phi''(x_i) |\phi''(x_f)|} e^{-\Delta E / \delta},
    \end{equation}
    where $\Delta E = \phi(x_f) - \phi(x_i)$ is the height of the potential barrier.
    The square root prefactor is proportional to the product of two characteristic frequencies, $\phi''(x_i)$ and $|\phi''(x_f)|$,
    associated with the curvature of the potential at the bottom $x_i$ of the well and at the top $x_f$ of the barrier, respectively.

    This framework can be generalized to $N$ dimensions,
    \begin{equation} \label{eq:main_dynamic}
        \dot{\x} = \f(\x) + \sqrt{2\delta}\et,
        \quad \et \overset{\text{iid}}{\sim} \mathcal{N}(0,\mathbf{I}),
    \end{equation}
    where the force becomes a generalized force that can include a solenoidal term in its Hodge decomposition
    (a generalization of the Helmholtz decomposition to higher dimensions) \cite{zhou2012quasi,glotz2023}.
    The \emph{generalized force} $f(x)$ can be expressed as the sum of a conservative (\emph{longitudinal})
    force derived from a scalar potential $\phi(x)$ and a non-conservative (\emph{transversal})
    force derived from an anti-symmetric (anti-Hermitian in the complex case) matrix $\A(x)$
    \begin{equation}	\label{eq:force_decomp}
        f_i(x) = -\partial_i \phi(x) + \sum_j \partial_j A_{ij}(x),
    \end{equation}
    where $\partial_j$ denotes the derivative with respect to the $j^{\text{th}}$-component of the state vector $\x$.
    In this case, the dynamics is said to be non-variational since it does not necessarily derive from a least action principle.

    Near a stable fixed point $\x_0$, the dynamics can be linearized
    \begin{equation} \label{eq:approx_dynamic}
        \dot{\x} \approx \mathbf{J}_f(\x_0)\x + \sqrt{2\delta}\et,
    \end{equation}
    where $\mathbf{J}_f(\x_0)$ is the Jacobian matrix of $\f(\x)$ at $\x_0$.
    Stability requires the real parts of all eigenvalues of $\mathbf{J}_f(\x_0)$ to be negative.

    Non-variational systems can also exhibit non-normality,
    characterized by non-orthogonal eigenvectors when  $[\mathbf{J}_f(\x_0), \mathbf{J}_f(\x_0)^\dag] \neq 0$.
    This property is common in dynamical systems,
    as non-normal matrices dominate the space of all matrices,
    and it can be quantified by the condition number $\kappa = \sigma_1 / \sigma_n$ \cite{Embree2005}
    (where $\sigma_1$ and $\sigma_n$ are the largest and smallest singular values of $\mathbf{J}_\f(\x_0)$, respectively).
    While normal systems have $\kappa = 1$,
    high non-normality ($\kappa \gg 1$) leads to transient amplification of fluctuations proportional to $\kappa^2$ \cite{troude2024}.
    This amplification arises from the interaction between two orthogonal components called the non-normal mode and its reaction mode.
    When a perturbation excites the non-normal mode,
    it triggers rapid growth in the reaction mode,
    amplifying the initial perturbation.

   Intuitively, one may anticipate that the amplification of fluctuations induced by non-normal dynamics significantly alters the escape behavior in the classical Kramers problem.
   We demonstrate below that this amplification mechanism effectively renormalizes the noise amplitude, thereby reshaping the system's probabilistic dynamics.
   As a result, while the escape behavior of non-normal systems remains structurally analogous to that of variational, normal systems,
   it operates as if under a renormalized temperature, scaled by the square of the system's condition number $\kappa$.

\section*{Effective Action and Transition Probability in Non-Normal Nonlinear Systems}

    In the previous section, we recalled the classical Kramers framework,
    where escape rates are determined by the interplay between a scalar potential $\phi(x)$
    and thermal noise of variance $\delta$.
    In higher dimensions, however, the generalized force \eqref{eq:force_decomp} admits
    a solenoidal (non-conservative) component.  

    In two dimensions, this decomposition can be written explicitly as
    \begin{equation}    \label{eq:dyn_2d}
        \begin{cases}
            \dot{x} = -\partial_x \phi(x,y) + \partial_y \psi(x,y) + \sqrt{2\delta}\,\eta_x, \\
            \dot{y} = -\partial_y \phi(x,y) - \partial_x \psi(x,y) + \sqrt{2\delta}\,\eta_y,
        \end{cases}
        \; \eta_x,\eta_y \overset{\text{i.i.d.}}{\sim} \mathcal{N}(0,1),
    \end{equation}
    where $\phi(x,y)$ is the scalar potential and $\psi(x,y)$ is the stream function generating the solenoidal flow.
    The conservative part determines the position of attractors and barriers,
    while the solenoidal part encodes rotational, circulation-like dynamics that are directly responsible for non-normal amplification.

    Within the large-deviation framework, the probability of observing a trajectory
    $\x(t) = (x(t),y(t))$ is weighted by the Onsager--Machlup action
    \begin{equation}
        S_\tau[x,y] = \frac{1}{4\delta} \int_0^\tau 
        \Big[ \dot{x} + \partial_x \phi - \partial_y \psi \Big]^2
        + \Big[ \dot{y} + \partial_y \phi + \partial_x \psi \Big]^2 \, dt .
    \end{equation}
    According to Freidlin--Wentzell theory, the transition probability distribution between
    $\x(0)=\abf$ and $\x(\tau)=\bb$ satisfies
    \begin{equation}
        \mathcal{P}^\delta \;\asymp\; \exp\!\left(-\frac{S}{\delta}\right),
        \qquad
        S := \inf_{\x(0)=\abf,\;\x(\tau)=\bb} S_\tau[\x],
    \end{equation}
    where the minimizing trajectory is the most probable escape path,
    also known as the instanton.

    To introduce non-normal behavior,
    we choose separable potentials of the form
    \begin{equation}
        \phi(x,y) = \phi_x(x) + \phi_y(y),
        \qquad
        \psi(x,y) = \kappa \psi_y(y) - \kappa^{-1}\psi_x(x),
    \end{equation}
    where the imbalance parameter $\kappa$ controls the strength of non-normality.
    For $\kappa>1$ (assumed throughout),
    perturbations in the $y$-direction trigger a non-normal response in the $x$-direction.
    We therefore identify $y$ as the non-normal mode and $x$ as its reaction \cite{troude2024}.

    Near a stable point $y\approx y^*$, with
    $\partial_y \phi_y(y^*)=\partial_y \psi_y(y^*)=0$,
    the dynamics of $y-y^*$ scale as $\kappa^{-1}$.
    Writing $y = y^* + \kappa^{-1}z + \mathcal{O}(\kappa^{-2})$,
    the leading-order term of the functional in $\kappa^{-1}$ becomes
    \begin{equation}
        S_\tau = \frac{1}{4}\int_0^\tau
        \Big[\dot{x} + \partial_x \phi_x(x) - \beta z\Big]^2 dt,
        \qquad
        \beta := \partial_y^2\psi_y(y^*).
    \end{equation}
    Optimizing over $z$ gives
    \begin{equation}
        z = \frac{1}{\beta}\left[\dot{x} + \partial_x \phi_x(x)\right].
    \end{equation}
    Substituting this result back into the lower-order terms in $\kappa^{-1}$ yields the minimized action functional
    (see \SM~for details)
    \begin{equation}
        S = \left(\frac{\kappa_c}{\kappa}\right)^2 S_{\text{eff}} + \mathcal{O}(\kappa^{-3}),
        \qquad
        S_{\text{eff}} = \min_{x} S_\text{eff}[x],
    \end{equation}
    with
    \begin{equation}    \label{eq:action_eff}
        \small
        S_{\text{eff}}[x] =
        \frac{1}{4\omega^2} \int_0^\tau
        \Big[\ddot{x} + (\omega + \partial_x^2\phi_x(x))\dot{x}
        + \omega\partial_x\phi_x(x) - \beta\partial_x\psi_x(x)\Big]^2 dt.
    \end{equation}
    Here $\omega = \partial_y^2\phi_y(y^*)$ is the mean-reversion rate,
    and $\kappa_c := \omega/\beta$ is a threshold setting the scale at which non-normal renormalization occurs.

    To leading order, the transition probability takes the form
    \begin{equation}
        \mathcal{P} \;\sim\; \exp\!\left(-\frac{S_\text{eff}}{\delta_\text{eff}}\right),
        \qquad
        \delta_\text{eff} = \delta\left(\frac{\kappa}{\kappa_c}\right)^2,
    \end{equation}
    revealing that the effective noise level is amplified by a factor $(\kappa/\kappa_c)^2$.

    Moreover, under the assumption of a fast mean-reversion rate ($\omega \gg 1$),
    the effective action functional \eqref{eq:action_eff} simplifies to
    \begin{subequations}
        \begin{align}
            &S_{\text{eff}}[x] \;\approx\;
            \frac{1}{4}\int_0^\tau \Big[\dot{x} + U_{\text{eff}}'(x)\Big]^2 dt, \\
            \text{where}\quad
            &U_{\text{eff}}(x) = \phi_x(x) - \frac{1}{\kappa_c}\psi_x(x) 
        \end{align}
    \end{subequations}
    defines an effective potential.
    Thus, the problem reduces to an overdamped Kramers problem with
    a modified potential and an amplified noise variance.

    A key insight of our framework is the distinction between the small input noise amplitude $\delta$
    (as assumed in classical Kramers theory) and the large output fluctuations induced by non-normality.
    The asymmetry of the dynamics, quantified by the non-normality index $\kappa$,
    amplifies the effective noise to
    \(\delta_{\text{eff}} = \delta(\kappa/\kappa_c)^2\).
    While fluctuation amplification is well documented in fields such as fluid dynamics and neuroscience
    \cite{Trefethen1993,Embree2005,ioannou1995nonnormality},
    our contribution is to provide a general analytical expression
    linking this phenomenon directly to transition probabilities in non-variational systems.

\section*{Effective Dynamics \& Numerical Application}

    To obtain a tractable approximation of the dynamics in the presence of strong non-normality,
    we exploit the separation of timescales induced by a fast mean-reversion in the $y$-direction \eqref{eq:dyn_2d}
    ($\omega \gg 1$).
    At leading order, the non-normal mode $y$ can be expressed as
    \begin{equation}
        y \;\approx\; \frac{1}{\omega}\left[\kappa^{-1}\partial_x \phi_x(x) + \sqrt{2\delta}\,\eta_y\right],
        \qquad \omega = \partial_y^2\phi_y(y^*).
        \label{fhbqgvq}
    \end{equation}
    Substituting this approximation into the dynamics of $x$ \eqref{eq:dyn_2d} gives the effective dynamics
    \begin{subequations}    \label{eq:eff_dyn}
        \begin{align}
            &\dot{x} \;=\; -U_\text{eff}'(x) + \sqrt{2\delta_\text{eff}}\,\eta, \\
            \text{where}\quad
            &U_\text{eff}(x) := \phi_x(x) - \frac{1}{\kappa_c}\psi_x(x), \\
            \text{and}\quad
            &\delta_\text{eff} := \delta\left[1 + \left(\frac{\kappa}{\kappa_c}\right)^2\right],
        \end{align}
    \end{subequations}
    and $\eta$ is a standard Gaussian white noise (see \SM).
    Thus, the effective dynamics of the reaction variable reduces to a
    one-dimensional overdamped Langevin equation,
    which has the same form as the classical Kramers problem, but with
    a renormalized potential $U_\text{eff}$ and an amplified noise variance $\delta_\text{eff}$.

    In this reduced description, the escape rate between a local minimum $x_i$ and the barrier top $x_f$
    is given by the Kramers formula
    \begin{equation}
        \Gamma \;\approx\; \frac{1}{2\pi}
        \sqrt{U_\text{eff}''(x_i)\,\big|U_\text{eff}''(x_f)\big|}
        \exp\!\left(-\frac{\Delta U_\text{eff}}{\delta_\text{eff}}\right),
        \label{wrthtn4i43}
    \end{equation}
    where $\Delta U_\text{eff} = U_\text{eff}(x_f)-U_\text{eff}(x_i)$.
    This explicit formula reveals that non-normality enters in two ways:
    through the modification of the potential $U_\text{eff}$,
    and through the amplification of the effective noise $\delta_\text{eff}$.
    
     For the large-deviation formalism underlying the standard solution of Kramers' problem to apply, 
    $\delta_\text{eff}$ must remain small compared to the energy barrier height. 
    This is the regime we consider in the illustrative treatment using (\ref{fhbqgvq}) and (\ref{eq:eff_dyn}), 
   aimed at providing a clear and intuitive understanding. Crucially, 
    $\delta_\text{eff}$ can be small yet still much larger than  $\delta$,
    allowing rate amplification to occur while the conditions for the derivation remain valid.

    Figure~\ref{fig:escape} compares the theoretical escape rate predicted by the renormalized Kramers formula (\ref{wrthtn4i43})
    with numerical simulations of the full two-dimensional stochastic dynamics,
    together with the standard deviation of the reaction variable $x$ as a function of $\kappa/\kappa_c$.
    The agreement confirms that the reduction faithfully captures
    the accelerated escape dynamics induced by non-normal amplification.

    Beyond the quantitative match of escape rates, the variance of $x$ also reveals
    a clear qualitative change in behavior.
    For $\kappa < \kappa_c$, the variance coincides with that of an
    Ornstein--Uhlenbeck process near a single stable equilibrium,
    consistent with Eq.~\eqref{eq:action_eff}.
    For $\kappa > \kappa_c$, the standard deviation approaches unity,
    reflecting frequent transitions between the two symmetric equilibria $x=\pm 1$.
    By symmetry, the probability of finding the system near $x=+1$ or $x=-1$
    is $1/2$ for each,
    so the stationary distribution is bimodal with variance exactly one.
    Hence, $\kappa_c$ acts as a critical threshold:
    below it, fluctuations remain localized;
    above it, the system explores both wells with equal probability,
    producing a phase-transition-like shift in the long-time statistics.

    In summary, eliminating the fast variable $y$ yields
    an effective one-dimensional Langevin description,
    with an effective potential $U_\text{eff}$ and an effective noise level
    $\delta_\text{eff}$ that both depend on $\kappa$ and $\kappa_c$.
    This reduced formulation provides an explicit escape rate formula
    and quantitative predictions,
    validated here by numerical experiments.
    It thus bridges the general large-deviation derivations
    with explicit test cases and simulations (see further details in the \SM).
    \newline

    \setcounter{figure}{0}
    \begin{figure}[t]
        \centering
        \includegraphics[width=0.5\textwidth]{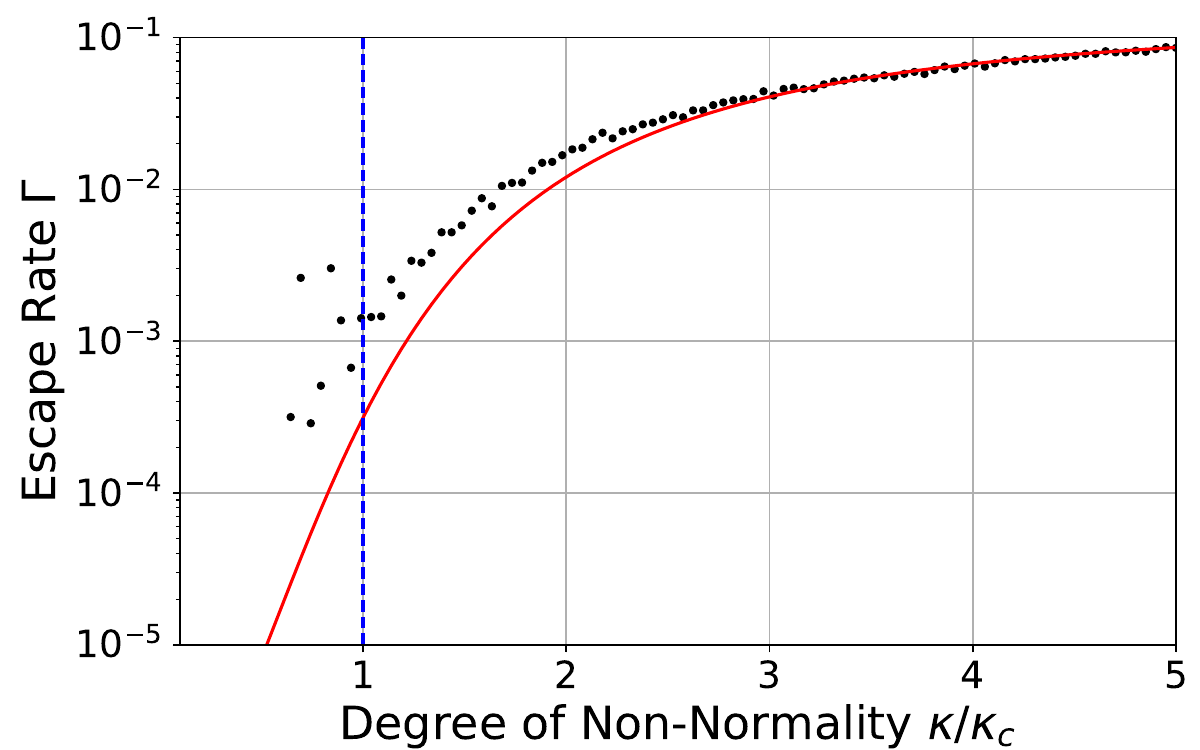}
        \includegraphics[width=0.5\textwidth]{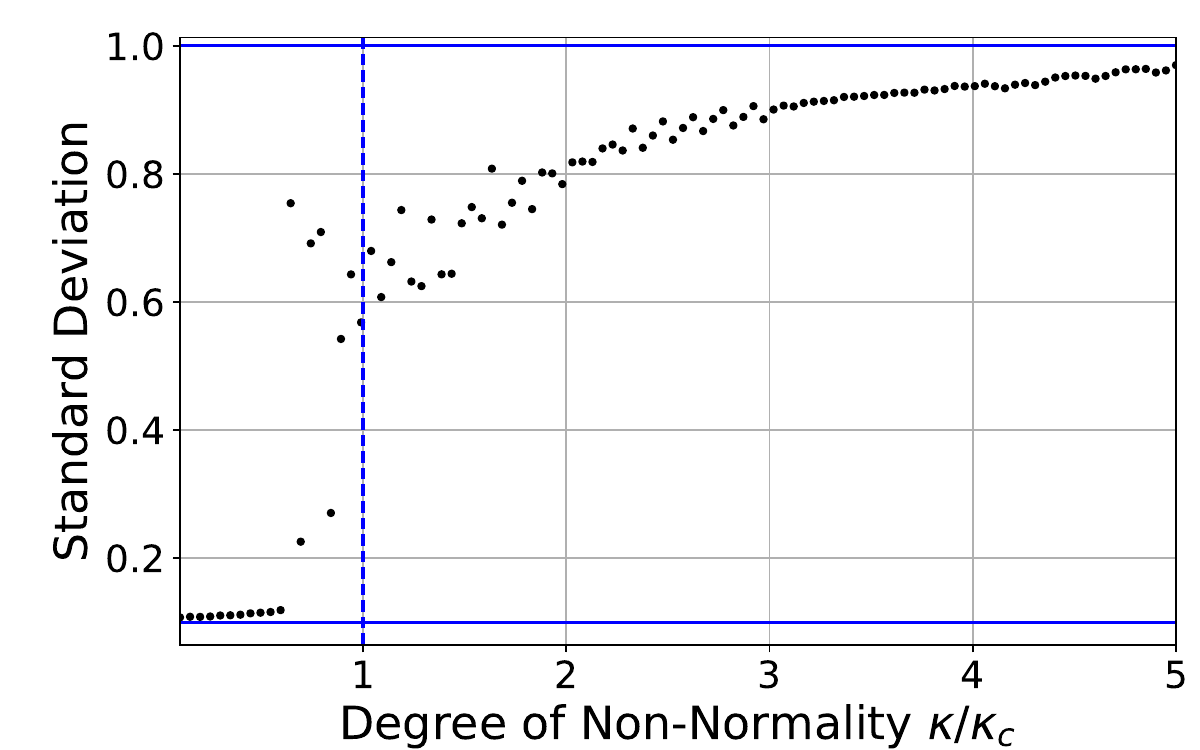}
        \caption{Comparison between theoretical predictions and numerical simulations
        of the dynamics \eqref{eq:numerical}. \\
        (\textbf{Top}) Escape rate $\Gamma$ as a function of $\kappa/\kappa_c$.
            black dots: numerical simulations of the full two-dimensional dynamics.
            Solid line: theoretical prediction from the renormalized Kramers formula
            $\Gamma \sim \exp\!\left(-\Delta U_\text{eff}/\delta_\text{eff}\right)$,
            with $U_\text{eff}$ and $\delta_\text{eff}$ defined in \eqref{eq:eff_dyn}. \\
        (\textbf{Bottom}) Standard deviation of the reaction variable $x$
            as a function of $\kappa/\kappa_c$.
            For $\kappa < \kappa_c$, the variance matches that of an
            Ornstein--Uhlenbeck process localized near a stable equilibrium.
            For $\kappa > \kappa_c$, the system transits between $x=\pm 1$
            with equal probability, leading to $\sqrt{\mathrm{Var}(x)} \approx 1$. \\
            We used $\omega=1$, $\delta=0.01$, and $\kappa_c=10$, 
            and simulations are performed for $\kappa$ ranging from $\kappa_c/10$ to $5\kappa_c$.
            Each simulation runs over a total time $T=10^{5}$ 
            with integration step $\Delta t=0.1$, 
            corresponding to $N=10^{6}$ data points. 
            The escape rate is estimated as $\Gamma = 1/\langle\tau\rangle$, 
            where $\langle\tau\rangle$ is the mean first-passage time from $x>0$ to $x<0$ (or vice versa).
        }
        \label{fig:escape}
    \end{figure}

\section*{Non-Normality-Induced Phase Transition}

    The classical notion of criticality in dynamical systems is tied to spectral changes:
    as a control parameter approaches a bifurcation point, one or more eigenvalues
    cross the imaginary axis, reducing stability and giving rise to large fluctuations.
    Recent work \cite{troude2025} has extended this picture to
    linear non-normal systems, introducing the concept of a critical degree of non-normality $\kappa_c$.
    Even when eigenvalues remain strictly stable, sufficiently large non-normality
    ($\kappa > \kappa_c$) induces pseudo-critical amplification,
    producing transient deviations that mimic the signatures of critical slowing down.

    Here, we extend this non-variational framework into the nonlinear regime,
    where the implications are even more profound.
    While in the linear case the system remains asymptotically stable for all $\kappa$,
    in the nonlinear stochastic setting, approaching $\kappa_c$
    destabilizes the vicinity of the equilibrium itself.
    The system escapes its local attractor along the reaction direction,
    driven not by a change in the deterministic spectrum but by
    the transient growth of fluctuations in the pseudo-critical phase.
    In other words, the equilibrium becomes effectively unstable
    once non-normality exceeds the critical threshold.

    This behavior is made explicit in Fig.~\ref{fig:escape},
    which shows a clear phase-transition-like change in the variance of the reaction variable $x$.
    For $\kappa < \kappa_c$, the system behaves as an Ornstein--Uhlenbeck process localized around a single well.
    For $\kappa > \kappa_c$, frequent transitions occur between the symmetric wells $x=\pm 1$,
    and the variance saturates at unity, corresponding to a bimodal distribution
    with equal occupation of both states.
    Thus, $\kappa_c$ plays the role of a critical  parameter:
    below it, fluctuations are confined; above it, global state switching becomes dominant.
  
    From the perspective of classical Kramers theory, this transition is highly non-traditional.
    Classically, for a system in contact with a fixed thermostat, 
    accelerated escape can only arise when the potential barrier is lowered,
    reducing $\Delta U$ in the exponential factor
    $\Gamma \sim \exp(-\Delta U / \delta)$.
    In contrast, in non-normal systems, the barrier remains fixed,
    but the effective noise level is rescaled to
    \(
        \delta_\text{eff} = \delta\left(\kappa/\kappa_c\right)^2
    \)
    via non-normality amplification of noise.
    Physically, if noise originates from thermal fluctuations,
    this corresponds to an effective temperature that can exceed
    the true temperature by more than an order of magnitude.
    For instance, in the simulations of Fig.~\ref{fig:escape},
    we fixed $\delta=0.01$ and varied $\kappa/\kappa_c$ from $0.1$ to $5$.
    At $\kappa/\kappa_c=5$, the effective noise is amplified by a factor of $26$,
    yielding $\delta_\text{eff}=0.26$.
    In this regime, the predicted escape rate increases from
    $\Gamma \sim 10^{-7}$ at $\kappa/\kappa_c=0.1$
    to $\Gamma \sim 0.1$ at $\kappa/\kappa_c=5$,
    in excellent agreement with simulations.
    While the WKB expansion underlying the Kramers approximation formally requires 
    $\delta_\text{eff} \ll 1$,  our numerics show it remains accurate even for $\delta_\text{eff}=0.26$, 
    indicating robustness beyond the strict asymptotic regime.
    
    The critical non-normality value $\kappa_c = \omega / \beta$ quantifies the balance between two competing 
    processes: the restorative tendency of the potential well, governed by the curvature 
    $\omega = \partial_y^2\phi_y(y^*)$, and the shear imposed by the non-normal coupling, set by $\beta := \partial_y^2\psi_y(y^*)$. 
    Much like the competition between interaction strength (here played by the restoring amplitude $\omega$) and thermal agitation
    (here played by the shear amplitude $\beta$)  that controls classical 
    phase transitions, this balance determines whether fluctuations are suppressed or 
    amplified. Physically, $\kappa_c$ acts as a threshold: for $\kappa < \kappa_c$, restoring 
    forces dominate and the system behaves as a conventional stable equilibrium, with noise 
    producing only minor perturbations. Once $\kappa > \kappa_c$, however, shear overwhelms 
    local curvature, noise is effectively multiplied, and escape rates increase dramatically. 
    In this sense, $\kappa_c$ defines the critical degree of non-normality beyond which the 
    system undergoes a qualitative change of regime, even though spectral stability is fully 
    preserved.
 
    We have therefore demonstrated a new type of phase transition:
    one controlled not by spectral criticality, but purely by non-normal amplification.
    This extends the concept of pseudo-criticality from the linear domain
    into the nonlinear stochastic regime,
    where it manifests as a qualitative change in global dynamics
    and an exponential acceleration of transition rates.
    By solving the Kramers problem in a non-variational, highly non-normal system,
    we reveal an amplification mechanism that effectively renormalizes temperature itself,
    opening a new perspective on rare-event dynamics in complex systems.

\section*{Application to DNA Methylation}

    DNA methylation is a key epigenetic mechanism controlling gene expression and cellular identity. 
    Its transitions are often observed on unexpectedly fast timescales,
    sometimes within minutes after environmental stimuli~\cite{busto2020stochastic}, 
    far exceeding predictions from classical variational Kramers-type models. 
    Several biological factors are known to facilitate rapid methylation,
    including chromatin accessibility~\cite{baubec2015genomic}, 
    transient metabolic fluctuations, and local feedback loops that propagate methylation marks across neighboring sites~\cite{day2010dna,kim2017dna}. 
    Yet, until now, a coherent theoretical framework capturing both bistability and rapid transitions has been lacking.

    We use DNA methylation as a particularly illustrative application to illustrate how non-normality provides such a framework \SM, 
    and as a roadmap for identifying when nonlinear systems can exhibit non-normal phase transitions. 
    Building on the bistable CpG dyad model of \cite{zagkos2019},
    which accounts for epigenetic memory through deterministic switching between unmethylated,
    hemimethylated, and fully methylated states, we show \SM~that the inclusion of stochasticity exposes structural limitations.
    Specifically, along the axis connecting the two stable equilibria, the dynamics lack a reaction mode compatible with non-normal amplification and instead reduce to Brownian-like motion,
    thereby preventing accelerated switching.
    We therefore introduce a minimal extension of the model that preserves the same stable and unstable equilibria, 
    while explicitly incorporating a degree of non-normality $\kappa$ into the force field.
    This modification aligns the reaction mode with the axis of bistability and allows transient deviations to be amplified, 
    thereby accelerating transitions between methylation states. 
    In this way, our framework reconciles the bistability necessary for epigenetic memory 
    with the rapid transition rates observed in vivo~\cite{busto2020stochastic}.

    The non-normality parameter $\kappa$ can be interpreted as an effective integrator of the enzymatic balance between DNA methyltransferases (DNMTs) and ten-eleven translocation (TET) dioxygenases.
    De novo methylation is catalyzed primarily by DNMT3A and DNMT3B, while DNMT1 ensures faithful propagation of methylation patterns during DNA replication. In contrast, TET enzymes initiate active demethylation by oxidizing 5-methylcytosine to successive oxidized derivatives, ultimately enabling replacement by unmodified cytosine.
    A relative increase in DNMT activity, or a decrease in TET activity, maps to higher values of $\kappa$, whereas enhanced TET activity or reduced DNMT activity corresponds to lower $\kappa$.
    The dynamic reversibility between methylation and demethylation can thus be represented as an asymmetry in the effective potential \SM.

    When $\kappa \gtrsim \kappa_c$, stochastic fluctuations are strongly amplified, enabling rapid switching between methylation states.
    Such a regime may arise from DNMT overexpression or TET downregulation in cancer~\cite{Klutstein2016,McGovern2012},
    or from efficient maintenance methylation, where DNMT1 rapidly restores marks after replication while demethylation remains weak~\cite{McGovern2012}.
    Numerical analysis \SM ~shows that in this regime the system tends to escape the hypomethylated equilibrium $z_{3,-}$ and becomes trapped in the hypermethylated state $z_{3,+}$,
    because the effective barrier separating $z_{3,-}$ from the unstable saddle is lower than the one protecting $z_{3,+}$.
    This asymmetry implies that transitions are not readily reversible: demethylation-prone dynamics emerge only under TET hyperactivity or reduced DNMT levels~\cite{Li2010,Hunter2012},
    and in aging contexts where impaired DNMT1 maintenance progressively erodes methylation patterns~\cite{Li2010}.

    In summary, DNA methylation illustrates how existing nonlinear models can be extended to account for both bistability and rapid stochastic transitions through non-normality. 
    Our results demonstrate that high-$\kappa$ regimes naturally explain the fast epigenetic responses seen in experiments~\cite{busto2020stochastic}, 
    and capture the gradual demethylation observed in development and aging. 
    Thus, DNA methylation exemplifies how non-normal phase transitions can emerge in biological systems, 
    providing a concrete application of our generalized non-variational Kramers framework.   

\section*{Discussion}

    We have presented a unified framework demonstrating that non-normality can act as a catalyst of phase transitions.  
    By analyzing escape dynamics in non-variational systems through the lens of large-deviation theory,  
    we showed that the transient amplification intrinsic to non-normal operators leads to a renormalization of noise amplitude,  
    which can be interpreted as an effective temperature.  
    This rescaling depends on a single control parameter, the non-normality index $\kappa$, and its critical threshold $\kappa_c$.  
    When $\kappa \gtrsim \kappa_c$, the system undergoes a qualitative change in its dynamics:  
    stable equilibria lose their practical relevance because fluctuations are amplified strongly enough to induce escapes,  
    even though spectral stability is fully preserved.  

    This result establishes a fundamentally new form of phase transition:  
    one that is not triggered by eigenvalues crossing zero, but by the geometry of the eigenvectors themselves.  
    In other words, pseudo-criticality~\cite{troude2024,troude2025} is not merely a mimicry of classical critical phenomena,  
    but can in fact constitute the origin of genuine transitions.  
    Within this perspective, non-normal systems expand the classical universality of phase transitions,  
    introducing a new pathway by which noise and structure interact to produce abrupt changes.  
   Whereas in the Kramers formalism the effect would be described as a mere lowering of the potential barrier, 
   in reality it stems from non-normal amplification, which boosts the effective noise level, sometimes by orders of magnitude, thereby enabling escape rates unattainable under ambient conditions.
   
    The broader implications are considerable.  
    In biology, we showed that DNA methylation,
    a cornerstone of epigenetic regulation,
    naturally falls into this class of dynamics.  
    By extending the bistable nonlinear model of CpG dyads~\cite{zagkos2019} to include explicit non-normality,  
    we have reconciled the coexistence of bistability (which secures epigenetic memory)  
    with fast stochastic switching (which enables rapid adaptation),  
    thereby explaining observed methylation dynamics on timescales of minutes~\cite{busto2020stochastic}.  
    More generally, the mechanism we describe can apply to diverse domains:  
    from abrupt climate tipping points driven by inherently non-variational fluid flows,  
    to critical transitions in ecosystems, markets, or engineered networks.  

    In summary, our work establishes that phase transitions need not be spectrally induced.  
    They can emerge purely from non-normal amplification,  
    through a control parameter $\kappa$ that redefines effective noise and reshapes escape dynamics.  
    This conceptual advance not only unifies disparate observations of sudden transitions across disciplines,  
    but also provides a predictive, compact analytical tool for quantifying them.  
    Non-normal catalysis thus offers a new universality class of transitions,  
    bridging stochastic theory, nonlinear dynamics, and real-world phenomena in physics, biology, and beyond.

{\bf Acknowledgements}:
We thank Sandro Lera and Ke Wu for stimulating discussions and constructive feedbacks on previous versions of this manuscript.
This work was partially supported by the National Natural Science Foundation of China (Grant No. T2350710802 and No. U2039202), Shenzhen Science and Technology Innovation Commission Project (Grants No. GJHZ20210705141805017 and No. K23405006), and the Center for Computational Science and Engineering at Southern University of Science and Technology.

\balance
\bibliographystyle{naturemag}  
\bibliography{bibliography} 


\clearpage
\onecolumn
\appendix

\section*{\Large\bf Supplementary Materials}
\setcounter{section}{0}
\renewcommand{\thesection}{S\arabic{section}}
\renewcommand{\thesubsection}{S\arabic{section}.\arabic{subsection}}

We first introduce the mathematical framework and key concepts used throughout the Supplementary Material (SM).
We provide the necessary background and definitions for understanding the derivations and analysis presented in the subsequent sections.

This Supplementary Material aims to present analytical derivations demonstrating how non-normality impacts the probability of transitions between states in non-variational systems.
The main results show that non-normality leads to an amplification mechanism,
enabling faster transitions between states and even causing systems to exit stable equilibrium more rapidly due to this amplification.

\tableofcontents

\newpage

\section{Mathematical Framework}

    In this appendix, we provide the mathematical background and assumptions underlying our analysis.
    We first introduce the dynamical system of interest,
    before reformulating the problem in the language of stochastic calculus.

    \subsection{Generalized Langevin System}

    We consider an overdamped Langevin dynamics describing the evolution of an $N$-dimensional state vector $\x$,
    subject to a generalized force $\f$ and stochastic fluctuations
    \begin{equation}    \label{eq:main_dynamic_apx}
        \dot{\x} = \f(\x) + \sqrt{2\delta}\,\et,
        \qquad \et \overset{\text{iid}}{\sim} \mathcal{N}(0,\mathbf{I}),
    \end{equation}
    where $\delta$ denotes the noise amplitude.
    In physical systems where the noise originates from thermal fluctuations, one has $\delta = k_B T$,
    with $k_B$ Boltzmann's constant and $T$ the temperature.

    By Hodge decomposition—a generalization of the Helmholtz decomposition to higher dimensions \cite{zhou2012quasi,glotz2023}
    -- the generalized force can be expressed as
    \begin{equation} \label{eq:force_decomp_apx}
        f_i(\x) = -\partial_i \phi(\x) + \sum_j \partial_j A_{ij}(\x),
    \end{equation}
    where $\phi(\x)$ is a scalar potential and $\A(\x)$ is an anti-symmetric (anti-Hermitian in the complex case) matrix.
    The first term represents a conservative (\emph{longitudinal}) force,
    while the second corresponds to a non-conservative (\emph{transversal}) force.
    Systems of this type are called \emph{non-variational},
    as they do not generally derive from a least-action principle.

    Near a stable fixed point $\x_0$, the dynamics can be linearized as
    \begin{equation} \label{eq:approx_dynamic_apx}
        \dot{\x} \approx \J_f(\x_0)\,\x + \sqrt{2\delta}\,\et,
    \end{equation}
    where $\J_f(\x_0)$ is the Jacobian of $\f$ at $\x_0$.
    Stability requires that all eigenvalues of $\J_f(\x_0)$ have negative real parts.

    If $\A=0$ in \eqref{eq:force_decomp_apx}, then the system is variational and $\J_f(\x_0)$ is Hermitian,
    i.e. $\J_f(\x_0) = \J_f(\x_0)^\dagger$.
    Our focus, however, is on \emph{non-normal systems}, characterized by
    \(
        [\J_f(\x_0), \J_f(\x_0)^\dagger] \;\neq\; 0,
    \)
    which implies that $\J_f(\x_0)$ cannot be diagonalized in a unitary basis.
    Even when such systems are linearly stable, transient deviations may be strongly amplified,
    and this amplification grows with the degree of non-normality.

    A natural quantitative measure of non-normality is the condition number $\kappa$ of the eigenbasis transformation of $\J_f(\x_0)$ \cite{troude2024}.
    In this work, we investigate the leading-order behavior of escape probabilities in the asymptotic regime $\kappa \gg 1$.

    \subsection{Problem Statement}

    We now reformulate the problem using the framework of stochastic calculus.
    Let $(\Omega,\mathcal{F},\mathbb{P})$ be a filtered probability space supporting an $N$-dimensional Brownian motion $\W$.
    We consider the Itô stochastic differential equation
    \begin{equation}    \label{eq:ito_sde_apx}
        d\x = \f(\x)\,dt + \sqrt{2\delta}\,d\W,
        \qquad \x\in\mathbb{R}^N,\;\; 0 < \delta \ll 1,
    \end{equation}
    where $\f \in C^2(\mathbb{R}^N;\mathbb{R}^N)$ is the drift function.
    We assume that $\f$ satisfies a sub-quadratic growth condition,
    ensuring the existence of global strong solutions.
    Equation \eqref{eq:ito_sde_apx} is equivalent to \eqref{eq:main_dynamic_apx},
    but expressed in stochastic calculus notation.

    We focus on two disjoint open subsets $A,B \subset \mathbb{R}^N$,
    each containing a hyperbolic equilibrium of $\f$: $\abf \in A$ and $\bb \in B$.
    The central object of interest is the transition probability
    \begin{equation}
        \mathcal{P}^\delta := \mathbb{P}^\delta_{\abf}\!\left\{\tau_B < \tau_{\partial A}\right\},
    \end{equation}
    where the stopping times are defined by
    \begin{equation}
        \tau_B := \inf_{t>0}\{\x_t \in B\}, 
        \qquad
        \tau_{\partial A} := \inf_{t>0}\{\x_t \notin A\}.
    \end{equation}
    We are particularly interested in the small-noise limit $\delta \to 0^+$.

    The main objective of this work is to quantify how strong non-normality of the Jacobian $\J(\x) := D\f(\x)$ modifies the asymptotic behavior of $\mathcal{P}^\delta$.

    \subsection{Large-Deviation Framework}

    For any fixed time horizon $\tau > 0$, the law of the process $(\x_t)_{t \in [0,\tau]}$ satisfies a Large-Deviation Principle (LDP) on the space $C([0,\tau];\mathbb{R}^N)$,
    characterized by the good rate function (action functional)
    \begin{equation}    \label{eq:func_action_apx}
        S_\tau[\x] = \frac{1}{4}\int_0^\tau \bigl\|\dot{\x}_t - \f(\x_t)\bigr\|^2 dt,
        \qquad \x \in H^1([0,\tau];\mathbb{R}^N).
    \end{equation}
    According to Freidlin--Wentzell theory, the transition probability admits the asymptotic representation
    \begin{equation}    \label{eq:prob_fw_apx}
        \mathcal{P}^\delta \;\asymp\; \exp\!\left(-\frac{S}{\delta}\right),
        \qquad 
        S := \inf_{\tau>0}\;\inf_{\x(0)=\abf,\;\x(\tau)\in B}\, S_\tau[\x],
    \end{equation}
    where $S$ is the \emph{quasi-potential} between the equilibrium $\abf$ and the set $B$.
    The second $\inf_{\x(0)=\abf,\;\x(\tau)\in B}$ in (\ref{eq:prob_fw_apx}) is taken over all 
    paths starting from $\x(0)$ and ending in domain $B$.
    
    To ensure the validity of the large-deviation approximation and the asymptotic expansions employed,
    we impose the following assumptions:
    \begin{itemize}
        \item \textbf{(A1) Smoothness:}
        The drift $\f$ is twice continuously differentiable ($\f \in C^2$) and satisfies a sub-quadratic growth bound,
        i.e.\ $\|\f(\x)\|\le C(1+\|\x\|)$.
        \item \textbf{(A2) Hyperbolicity:}
        The Jacobians $D\f(\abf)$ and $D\f(\bb)$ are hyperbolic, meaning that all their eigenvalues have strictly negative real parts.
        \item \textbf{(A3) Unique Minimizer:}
        The optimization problem in \eqref{eq:prob_fw_apx} admits a unique minimizing path $\x^\ast$.
        \item \textbf{(A4) Local Uniformity:}
        The asymptotic expansions derived later remain uniformly valid in the scaling parameter $\kappa$ within a local neighborhood of interest.
    \end{itemize}
    Under \textbf{(A1)--(A3)}, the LDP and associated saddle-point approximations are rigorous.
    Assumption \textbf{(A4)} ensures uniformity in the large shear introduced by non-normality,
    as measured by $\kappa$.
    \newline

    Different approaches can be used to evaluate the quasi-potential $S$.
    One possibility is to identify the \emph{instanton} path by minimizing the Lagrangian
    \begin{equation}
        L(\dot{\x},\x) := \tfrac{1}{4}\bigl\|\dot{\x} - \f(\x)\bigr\|^2,
    \end{equation}
    which leads to the Euler--Lagrange equation
    \begin{equation}
        \frac{d}{dt}\frac{\partial L}{\partial \dot{\x}} \;=\; \frac{\partial L}{\partial \x}.
    \end{equation}
    Alternatively, and more conveniently for our analysis,
    one may adopt the Hamilton--Jacobi formalism.
    Assuming the minimizing action can be represented as a scalar field $S(\x)$,
    for each infinitesimal time step $\delta t$, the optimal path moves with velocity $\vv$ and continues optimally thereafter from the updated position $\x - \vv\,\delta t$
    \begin{align}
        S(\x) 
        &= \inf_{\vv \in \mathbb{R}^N}\Bigl\{ S(\x - \vv \delta t) + L(\vv,\x)\,\delta t + \mathcal{O}(\delta t^2)\Bigr\} \\
        &= \inf_{\vv \in \mathbb{R}^N}\Bigl\{ S(\x) - \vv \cdot \nabla_\x S(\x)\,\delta t + L(\vv,\x)\,\delta t + \mathcal{O}(\delta t^2)\Bigr\}, \\
        \Rightarrow\quad
        0 &= \inf_{\vv \in \mathbb{R}^N}\Bigl\{ L(\vv,\x) - \vv \cdot \nabla_\x S(\x) + \mathcal{O}(\delta t)\Bigr\}.
    \end{align}
    Canceling $S(\x)$ and retaining terms of order $\delta t$,
    the Hamiltonian is obtained via a Legendre transform
    \begin{align}
        H(\x,\p) &:= \sup_{\vv}\Bigl\{ \p \cdot \vv - L(\vv,\x) \Bigr\}, \qquad \p = \nabla_\x S(\x), \\
                &= \p \cdot \f(\x) + \|\p\|^2.
    \end{align}
    The minimizing action $S(\x)$ is therefore characterized by the Hamilton--Jacobi PDE
    \begin{equation}    \label{eq:hj_apx}
        H(\x,\nabla_\x S) \;=\; \|\nabla_\x S(\x)\|^2 + \f(\x)\cdot\nabla_\x S(\x) = 0.
    \end{equation}

    This equation is equivalent to the leading-order stationary Fokker--Planck equation
    \begin{equation}
        \delta\,\Delta_\x P(\x) - \nabla_\x \cdot \bigl[\f(\x)P(\x)\bigr] = 0,
    \end{equation}
    where $P(\x)$ is the stationary density.
    With the ansatz $P(\x) = e^{-S(\x)/\delta}$, the PDE becomes
    \begin{equation}    \label{eq:fp_apx}
        \|\nabla_\x S(\x)\|^2 + \f(\x)\cdot\nabla_\x S(\x)
        \;=\; \delta \bigl[\Delta_\x S(\x) + \nabla_\x \cdot \f(\x)\bigr].
    \end{equation}
    In the small-noise limit $\delta \to 0^+$,
    solutions of \eqref{eq:fp_apx} converge to the Hamilton--Jacobi solution \eqref{eq:hj_apx}.
    \newline

    We conclude that, in systems governed by an LDP,
    the central task is to extract the leading-order contribution to the quasi-potential $S$,
    which directly determines the exponential scaling of escape probabilities and transition rates.

    \subsection{Summary \& Main Statement}

    We have now formalized the dynamical system of interest \eqref{eq:main_dynamic_apx},
    and expressed the problem of estimating escape probabilities in terms of the quasi-potential $S$,
    as a function of the degree of non-normality $\kappa$.
    When $\kappa=1$, the system is normal; in the limit $\kappa \to \infty$,
    the system is ``highly'' non-normal.
    The LDP framework highlights that the key objective is to determine the leading-order dependence of $S$ on $\kappa$.

    \noindent\textbf{Proposition.} 
    \textit{
        In the limit of a highly non-normal system,
        i.e. $\kappa \gg \kappa_c$ where $\kappa_c$ is a critical threshold beyond which non-normal effects dominate the dynamics,
        the quasi-potential can be expressed as
        \begin{equation}
            S \;=\; \left(\frac{\kappa_c}{\kappa}\right)^2 S_{\text{eff}} \;+\; \mathcal{O}\!\left(\left(\tfrac{\kappa_c}{\kappa}\right)^3\right).
        \end{equation}
        At leading order in $\kappa/\kappa_c$,
        the exponent in \eqref{eq:prob_fw_apx} therefore simplifies to
        \begin{equation}
            \frac{S}{\delta} \;\approx\; \frac{S_{\text{eff}}}{\delta_{\text{eff}}}, 
            \qquad \delta_{\text{eff}} \sim \kappa^2 \delta,
        \end{equation}
        showing that non-normality effectively renormalizes the noise scale in Kramers-type problems,
        producing a significant amplification of escape rates.
    }

    The derivation of this proposition is given below.
    This result extends the framework of \cite{troude2025},
    which established a unifying description of amplification mechanisms in non-normal \emph{linear} systems,
    to the more general case of non-normal \emph{nonlinear} systems subject to Gaussian (thermal) fluctuations.

    The parameter $\kappa_c = \omega / \beta$, where $ \omega := \left.\frac{\partial^2 \phi_y}{\partial y^2}\right|_{y=y^*}$ (\ref{th3ybgqfvq}) and
    $\beta := \left.\frac{\partial^2 \psi_y}{\partial y^2}\right|_{y=y^*}$ (\ref{thunghb34})
    quantifies the balance between two competing 
    processes: the restorative tendency of the potential well, governed by the curvature 
    $\omega$, and the shear imposed by the non-normal coupling, set by $\beta$. 
    Physically, $\kappa_c$ acts as a threshold separating regimes where fluctuations are 
    either suppressed or strongly amplified. For $\kappa < \kappa_c$, restoring forces 
    dominate and the system behaves like a conventional stable equilibrium, with noise 
    producing only small perturbations. Once $\kappa$ exceeds $\kappa_c$, however, the shear 
    overwhelms the local curvature, so that noise is effectively multiplied and escape rates 
    increase dramatically. In this sense, $\kappa_c$ marks the critical degree of non-normality 
    beyond which the system undergoes a qualitative change of regime, despite its eigenvalues 
    remaining stable.

\section{Non-Normal Amplification of Stochastic Noise}

    Amplification of stochastic noise in \emph{linear} non-normal systems has been studied extensively in the past \cite{Biancalani2017,troude2024,troude2025}. 
    The interplay between non-normality and nonlinearity has also been noted in hydrodynamic contexts \cite{FarrellIoannou1996a}. 

    The purpose of this section is to demonstrate that, in the limit of a ``highly'' non-normal system,
    the noise variance rescaled by the factor  $\kappa^2$. 
    In particular, assuming the existence of a unique non-normal mode \cite{troude2024},
    the dynamics can be reduced to two dimensions:
    one associated with the non-normal mode itself, and the other with its reaction mode. 
    In this reduced setting, the matrix potential $\A$ from \eqref{eq:force_decomp_apx} can be written as
    \begin{equation}
        \A(\x) = \Q \psi(\x),
        \qquad \Q =
        \begin{pmatrix}
            0 & 1 \\
            -1 & 0
        \end{pmatrix},
        \qquad 
        \psi(\x) = \kappa\,\psi_y(y) - \kappa^{-1}\psi_x(x).
    \end{equation}
    Note that the scalar potential $\psi(\x)$ is separable in $x$ and $y$.
    We also assume a separable scalar potential for $\phi$,
    i.e.\ $\phi(\x) = \phi_x(x) + \phi_y(y)$.
    The generalized force \eqref{eq:force_decomp_apx} then takes the form
    \begin{equation}    \label{eq:force_2d_apx}
        \f(\x) = -\nabla\phi(\x) + \Q \nabla\psi(\x).
    \end{equation}

    Accordingly, the Jacobian of $\f$ at each point $\x=(x,y)$ is given by
    \begin{equation}
        \J(\x) := D\f(\x) = 
        \begin{pmatrix}
            -\partial_x^2 \phi_x(x) & \kappa\,\partial_y \psi_y(y) \\
            \kappa^{-1}\partial_x^2 \psi_x(x) & -\partial_y^2 \phi_y(y)
        \end{pmatrix}.
    \end{equation}
    The corresponding eigenvalues are
    \begin{equation}
        \lambda_\pm(\x) 
        = -\tfrac{1}{2}\bigl(\partial_x^2\phi_x(x) + \partial_y^2\phi_y(y)\bigr)
        \;\pm\; \tfrac{1}{2}\sqrt{ \bigl(\partial_x^2\phi_x(x) - \partial_y^2\phi_y(y)\bigr)^2
        + 4\,\partial_y^2 \psi_y(y)\,\partial_x^2 \psi_x(x)}.
    \end{equation}
    Remarkably, the degree of non-normality $\kappa$ does not appear in the spectrum. 
    Thus, we obtain a reduced two-dimensional nonlinear system in which the stability is governed solely by the potentials $\phi_i$ and $\psi_i$ ($i = x,y$),
    while non-normality manifests exclusively through a shear controlled by $\kappa$. 
    As $\kappa$ increases, the magnitude of this shear grows.

    Our objective in what follows is to analyze how the quasi-potential $S$ depends on $\kappa$,
    and thereby deduce the scaling of transition rates with the degree of non-normality.

    \subsection{Effective Quasi-Potential}
    \label{apx:effective_potential}

    For the system defined by \eqref{eq:main_dynamic_apx} with the force field given by \eqref{eq:force_2d_apx},
    the action functional \eqref{eq:func_action_apx} can be decomposed as
    \begin{subequations}
    \begin{align}
        S_\tau[\x] &= S^x_\tau[x,y] + S^y_\tau[x,y], \\
        S^x_\tau[x,y] &= \frac{1}{4}\int_0^\tau \bigl\|\dot{x} + \partial_x \phi_x(x) - \kappa\,\partial_y \psi_y(y)\bigr\|^2 dt, \\
        S^y_\tau[x,y] &= \frac{1}{4}\int_0^\tau \bigl\|\dot{y} + \partial_y \phi_y(y) - \kappa^{-1}\partial_x \psi_x(x)\bigr\|^2 dt.
    \end{align}
    \end{subequations}
    To minimize the action \eqref{eq:func_action_apx},
    the force $\f(\x)$ must remain of order $\mathcal{O}(1)$ along the optimal trajectory.  
    This requires the non-normal shear contribution $\kappa\,\partial_y \psi_y(y)$ to remain at most $\mathcal{O}(1)$.  

    In the limit $\kappa \to \infty$, the dynamics along the non-normal mode can be expanded near an equilibrium point $y^*$ as
    \begin{equation}    \label{eq:approx_dynamic_y_apx}
        y = y^* + \kappa^{-1} z + \mathcal{O}(\kappa^{-2}),
        \qquad \text{with } \; \left.\partial_y \phi_y\right|_{y=y^*} = 0.
    \end{equation}
    Expanding the non-normal shear term in powers of $\kappa^{-1}$ yields
    \begin{equation}    \label{eq:approx_shear_apx}
        \kappa\,\partial_y \psi_y(y) 
        = \kappa\,\partial_y \psi_y(y^*) 
        + \partial_y^2 \psi_y(y^*)\,z + \mathcal{O}(\kappa^{-1}).
    \end{equation}
    Since $\kappa\,\partial_y \psi_y(y^*)$ must remain at most $\mathcal{O}(1)$,
    $y^*$ must also be a zero of $\partial_y \psi_y$.  
    Thus, the action functionals simplify to
    \begin{subequations}
    \begin{align}
        S^x_\tau[x,y] &= \frac{1}{4}\int_0^\tau 
        \bigl\|\dot{x} + \partial_x \phi_x(x) - \beta z + \mathcal{O}(\kappa^{-1})\bigr\|^2 dt,   \label{thunghb34}
        &\beta := \left.\partial_y^2 \psi_y\right|_{y=y^*}, \\
        S^y_\tau[x,y] &= \kappa^{-2} S^z_\tau[x,z], \\
        S^z_\tau[x,z] &= \frac{1}{4}\int_0^\tau 
        \bigl\|\dot{z} + \omega z - \partial_x \psi_x(x) + \mathcal{O}(\kappa^{-1})\bigr\|^2 dt,
        &\omega := \left.\partial_y^2 \phi_y\right|_{y=y^*}.
        \label{th3ybgqfvq}
    \end{align}
    \end{subequations}
    Since $S^y_\tau[x,y] = \mathcal{O}(\kappa^{-2})$ while $S^x_\tau[x,y] = \mathcal{O}(1)$,
    minimizing the action with respect to $z$ gives, up to $\mathcal{O}(1)$,
    \begin{equation}
        \dot{x} + \partial_x \phi_x(x) - \beta z + \mathcal{O}(\kappa^{-1}) = 0
        \quad \Rightarrow \quad
        z = \frac{1}{\beta}\bigl[\dot{x} + \partial_x \phi_x(x)\bigr] + \mathcal{O}(\kappa^{-1}).
    \end{equation}
    This cancels the leading-order contribution from $S^x_\tau$, leaving the reduced problem
    \begin{equation}
        S_\tau[\x] = \kappa^{-2} S^z_\tau[x,z] + \mathcal{O}(\kappa^{-3}),
        \qquad \text{subject to } \dot{x} + \partial_x \phi_x(x) = \beta z.
    \end{equation}
    Substituting this constraint into the reduced action, and using
    \begin{equation}
        \dot{z} = \tfrac{1}{\beta}\bigl[\ddot{x} + \partial_x^2 \phi_x(x)\,\dot{x}\bigr],
    \end{equation}
    we obtain the effective action functional
    \begin{equation}    \label{eq:effective_action_apx}
        S^{\mathrm{eff}}_\tau[x] 
        = \frac{1}{4\beta^2}\int_0^\tau 
        \bigl[\ddot{x} + (\partial_x^2 \phi_x(x) + \omega)\dot{x} 
        + \omega\,\partial_x \phi_x(x) - \beta\,\partial_x \psi_x(x)\bigr]^2 dt.
    \end{equation}
    Hence, the quasi-potential takes the asymptotic form
    \begin{equation}
        S = \kappa^{-2} S_{\mathrm{eff}} + \mathcal{O}(\kappa^{-3}).
    \end{equation}
    \newline

    In summary, the quasi-potential $S$ admits an effective one-dimensional representation,
    where the impact of non-normality appears solely as a $\kappa^{-2}$ scaling.  
    Thus, in the highly non-normal regime, the escape problem is reduced to an effective single-mode description,
    with $\kappa$ controlling the rescaling of the noise amplification.

    \subsection{Leading-Order Dynamics}

    In the previous section, we established that the leading-order behavior of the quasi-potential $S$ scales as $\kappa^{-2}$.  
    This follows from the observation that, at leading order, the influence of $\dot{y}$ vanishes,
    so that the action functional contributes only at order $\kappa^{-2}$.

    Here, we derive the effective leading-order dynamics.  
    We begin by expanding the dynamics of $y$ around its equilibrium $y^*$
    \begin{equation}
        \dot{y} = -\partial_y \phi_y(y) + \kappa^{-1}\partial_x \psi_x(x) + \sqrt{2\delta}\,\eta_y.
    \end{equation}
    Introducing the rescaled coordinate $z$ as in \eqref{eq:approx_dynamic_y_apx}, we obtain
    \begin{equation}
        \dot{z} = -\omega z + \partial_x \psi_x(x) + \kappa \sqrt{2\delta}\,\eta_y + \mathcal{O}(\kappa^{-1}),
    \end{equation}
    where $\omega := \partial_y^2 \phi_y(y^*)$.  

    In the fast-recovery regime $\omega \gg 1$, this reduces to
    \begin{equation}
        z = \frac{1}{\omega}\,\partial_x \psi_x(x) 
        + \frac{\kappa}{\omega}\sqrt{2\delta}\,\eta_y 
        + \mathcal{O}(\kappa^{-1}).
    \end{equation}
    Substituting into the dynamics of $x$ yields
    \begin{equation}
        \dot{x} = -\partial_x \phi_x(x) 
                + \frac{\beta}{\omega}\,\partial_x \psi_x(x) 
                + \frac{\kappa \beta}{\omega}\sqrt{2\delta}\,\eta_y 
                + \sqrt{2\delta}\,\eta_x 
                + \mathcal{O}(\kappa^{-1}),
    \end{equation}
    with $\beta := \partial_y^2 \psi_y(y^*)$.  

    Neglecting $\mathcal{O}(\kappa^{-1})$ terms, the effective leading-order dynamics becomes
    \begin{equation}
        \dot{x} = -\partial_x \phi_x(x) 
                + \frac{1}{\kappa_c}\,\partial_x \psi_x(x) 
                + \sqrt{2\delta_{\mathrm{eff}}}\,\eta_x,
        \qquad \delta_{\mathrm{eff}} = \delta\!\left(1 + \left(\tfrac{\kappa}{\kappa_c}\right)^2\right),
        \qquad \kappa_c = \frac{\omega}{\beta}.
    \end{equation}

    This result is consistent with the effective action derived in \eqref{eq:effective_action_apx},  
    valid in the regime $\kappa \gg \kappa_c$ and under the assumption of fast mean reversion along the $z$ direction,  
    i.e.\ $\omega \gg |\partial_x^2 \phi_x(x)|$.  
    In this limit, the quasi-potential takes the form
    \begin{equation}    \label{eq:leading_order_dynamic_apx}
        S[\x] \approx \frac{1}{4}\left(\frac{\kappa_c}{\kappa}\right)^2
        \int_0^\tau 
            \Bigl[\dot{x} + \partial_x \phi_x(x) - \tfrac{1}{\kappa_c}\partial_x \psi_x(x)\Bigr]^2 dt.
    \end{equation}
    \newline

    Thus, by invoking the fast-recovery approximation for the non-normal mode,  
    we recover a rescaling of the noise amplitude $\delta$ by a factor of $\kappa^2$.  
    Consequently, the quasi-potential $S$ is rescaled by $\kappa^{-2}$,  
    which captures the leading-order behavior in the limit $\kappa \gg \kappa_c$.

    \subsection{Hamilton--Jacobi}

    In the previous section, we showed that the leading-order behavior of the quasi-potential $S$ scales as $\kappa^{-2}$,  
    since at this order the influence of $\dot{y}$ vanishes,
    and the action functional contributes only at order $\kappa^{-2}$.

    We now recover the same scaling using the Hamilton--Jacobi framework \eqref{eq:hj_apx}.  
    As before, we assume that $\phi_y$ and $\psi_y$ share the same equilibrium $y^*$,
    and employ the expansions in \eqref{eq:approx_dynamic_apx} and \eqref{eq:approx_shear_apx}.  
    Under these assumptions, the Hamilton--Jacobi equation \eqref{eq:hj_apx} becomes
    \begin{subequations}
        \begin{align}
            &(\partial_y S)^2 + (\partial_x S)^2 
            + \bigl(\kappa\,\partial_y \psi_y - \partial_x \phi_x\bigr)\,\partial_x S 
            + \bigl(\kappa^{-1}\partial_x \psi_x - \partial_y \phi_y\bigr)\,\partial_y S = 0, \\
            \Rightarrow \quad
            &\kappa^2(\partial_z S)^2 + (\partial_x S)^2 
            + \bigl(\beta z - \partial_x \phi_x(x) + \mathcal{O}(\kappa^{-1})\bigr)\,\partial_x S \notag\\
            &\hspace{2cm} + \kappa^{-1}\bigl(\partial_x \psi_x - \omega z + \mathcal{O}(\kappa^{-1})\bigr)(\kappa\,\partial_z S) = 0,
            \qquad \text{since } \partial_y = \kappa\,\partial_z,
        \end{align}
    \end{subequations}
    with $\beta := \partial_y^2 \psi_y(y^*)$ and $\omega := \partial_y^2 \phi_y(y^*)$.
    \newline

    \noindent\textbf{Asymptotic expansion.}  
    Note that $\kappa\,\partial_z S$ is not $\mathcal{O}(\kappa)$ but $\mathcal{O}(1)$,
    since if $S(x,y)$ is smooth in $y$ and $y = y^* + \kappa^{-1} z$, then
    \begin{equation}
        S(x,y) = \sum_{n=0}^\infty S^{(n)}(x)\left(\frac{z}{\kappa}\right)^n,
        \qquad S^{(n)}(x) = \frac{1}{n!}\left.\frac{\partial^n S}{\partial y^n}\right|_{y=y^*}.
    \end{equation}
    Hence
    \begin{equation}
        \kappa\,\partial_z S = \sum_{n=0}^\infty (n+1)\,S^{(n+1)}(x)\left(\frac{z}{\kappa}\right)^n.
    \end{equation}

    Expanding the Hamilton--Jacobi equation in powers of $z/\kappa$, the $\mathcal{O}(1)$ term yields
    \begin{equation}
        \bigl(S^{(1)}(x)\bigr)^2 + \bigl(\partial_x S^{(0)}(x)\bigr)^2 
        + \bigl(\beta z - \partial_x \phi_x(x)\bigr)\,\partial_x S^{(0)}(x) = 0.
    \end{equation}
    Since the coefficients $S^{(n)}(x)$ are independent of $z$, this forces $\partial_x S^{(0)}(x) = 0$,
    so that $S^{(0)}$ is constant, and also $S^{(1)}(x) = 0$.  
    Thus, the leading-order structure of $S$ is
    \begin{equation}
        S(x,y) = S^{(0)} + \kappa^{-2} S^{(2)}(x)\,z^2 + \mathcal{O}(\kappa^{-3}).
    \end{equation}
    \newline

    \noindent\textbf{Next order.}  
    At $\mathcal{O}(\kappa^{-2})$, the Hamilton--Jacobi equation becomes
    \begin{equation}
        4 z^2 \bigl(S^{(2)}(x)\bigr)^2 
        + \bigl(\beta z - \partial_x \phi_x(x)\bigr)\,z^2\,\partial_x S^{(2)}(x) 
        + \bigl(\partial_x \psi_x(x) - \omega z\bigr)\,z\,S^{(2)}(x) = 0.
    \end{equation}
    Since $S^{(2)}(x)$ is independent of $z$, one would naively set $\partial_x S^{(2)}(x)=0$, implying $S^{(2)}(x)=0$.  
    However, this contradicts the previous results.  
    To resolve this, we incorporate the trajectory constraint
    \begin{equation}
        z = \frac{1}{\beta}\,\partial_x \phi_x(x).
    \end{equation}
    Substituting yields
    \begin{equation}
        z\,S^{(2)}(x) = \tfrac{1}{2}\left(\tfrac{\omega}{\beta}\,\partial_x \phi_x(x) - \partial_x \psi_x(x)\right).
    \end{equation}
    \newline

    \noindent\textbf{First-order solution.}  
    The Hamilton--Jacobi solution then reads
    \begin{equation}
        S(x,y) = S^{(0)} + \frac{1}{2}\left(\frac{\kappa_c}{\kappa}\right)^2
        \frac{\partial_x \phi_x(x)}{\omega}\left[\partial_x \phi_x(x) - \frac{1}{\kappa_c}\partial_x \psi_x(x)\right],
    \end{equation}
    where $\kappa_c = \omega/\beta$.

    This expression is fully consistent with the previous derivations:  
    the leading-order correction scales as $(\kappa_c/\kappa)^2$,
    and involves the same combination of potentials $\partial_x \phi_x(x) - \tfrac{1}{\kappa_c}\partial_x \psi_x(x)$
    as in \eqref{eq:effective_action_apx} and \eqref{eq:leading_order_dynamic_apx}.

    \subsection{Validity of the WKB Approximation}

    In the previous sections, we have shown three complementary methods leading to the same conclusion:  
    the quasi-potential $S$ admits a leading-order term proportional to $(\kappa_c/\kappa)^2$.  
    Equivalently, this scaling corresponds to a rescaling of the noise amplitude by a factor $(\kappa/\kappa_c)^2$ in the limit $\kappa \gg \kappa_c$.  
    Thus, the effective noise level is given by
    \(
        \delta_{\text{eff}} \sim \kappa^2 \delta.
    \)
    However, this derivation relied on a WKB approximation,
    which assumes that the noise intensity $\delta$ is sufficiently small.  
    To justify this approximation, we show that the amplification of $\delta$ by $\kappa^2$ occurs only at leading order,
    and not in the full expansion.
    \newline

    \noindent\textbf{Two-scale expansion.}  
    From the Fokker--Planck equation \eqref{eq:fp_apx},
    we expand the quasi-potential in powers of $\delta$
    \begin{equation}
        S(x,y) = \sum_{n=0}^\infty \delta^n S_n(x,y).
    \end{equation}
    At each order we obtain
    \begin{subequations}
    \begin{align}
        \mathcal{O}(1): \quad 
            & \|\nabla_\x S_0(\x)\|^2 + \f(\x)\cdot\nabla_\x S_0(\x) = 0, \\
        \mathcal{O}(\delta): \quad 
            & 2 \nabla_\x S_0(\x)\cdot\nabla_\x S_{1}(\x) + \f(\x)\cdot\nabla_\x S_1(\x)
            = \Delta_\x S_{0}(\x) + \nabla_\x\cdot\f(\x), \\
        \mathcal{O}(\delta^n): \quad 
            & \sum_{k=0}^n \nabla_\x S_k(\x)\cdot\nabla_\x S_{n-k}(\x) + \f(\x)\cdot\nabla_\x S_n(\x) 
            = \Delta_\x S_{n-1}(\x), 
            \quad n\geq 2 .
    \end{align}
    \end{subequations}
    For each coefficient $S_n(x,y)$, we introduce a second expansion in $1/\kappa$
    \begin{equation}
        S_n(x,y) = \sum_{m=0}^\infty \kappa^{-m} S_n^{(m)}(x,y).
    \end{equation}
    \newline

    \noindent\textbf{Conclusion.}  
    The scaling $\delta_{\text{eff}} \sim \kappa^2 \delta$ emerges only at the leading order in both 
     in both $\delta$ and $1/\kappa$. Higher-order corrections in the two-scale expansion do not introduce additional amplifications of this type, 
     but instead contribute subdominant terms. This shows that the apparent growth of the effective noise level with $\kappa$
     reflects a leading-order renormalization rather than a breakdown of the asymptotics. 
     Consequently, the WKB approximation remains internally consistent: although $\delta_{\text{eff}}$ may become significantly larger than the bare noise 
    $\delta$, the expansion is still controlled by the small parameter $\delta$, ensuring the validity of the quasi-potential analysis and of the 
    WKB approximation in this regime.
    
    \subsection{Kramers Escape Rate}

    We now focus on the escape rate of the system.  
    Neglecting terms of order $\mathcal{O}(\kappa^{-1})$,
    the two-dimensional dynamics in $(x,y)$ reduce to an effective one-dimensional dynamics in $x$
    \begin{subequations}    \label{eq:dyn_eff_apx}
    \begin{align}
        \dot{x} &= -U_{\text{eff}}'(x) + \sqrt{2\delta_{\text{eff}}}\,\eta, \\
        U_{\text{eff}}'(x) &= \partial_x\phi_x(x) - \frac{1}{\kappa_c}\partial_x\psi_x(x), 
        \qquad 
        \delta_{\text{eff}} = \delta\left(1 + \left(\frac{\kappa}{\kappa_c}\right)^2\right).
        \label{eq:deltaeff}
    \end{align}
    \end{subequations}

    If $x_i$ denotes a stable equilibrium and $x_f$ the corresponding unstable point,  
    the Kramers escape rate is
    \begin{equation}    \label{eq:escape_apx}
        \Gamma = \frac{1}{2\pi}
        \sqrt{U_{\text{eff}}''(x_i)\,\bigl|U_{\text{eff}}''(x_f)\bigr|}
        \exp\!\left[-\frac{\Delta E_{\text{eff}}}{2\delta_{\text{eff}}}\right],
        \quad\text{where }
        \Delta E_{\text{eff}} = U_{\text{eff}}(x_f) - U_{\text{eff}}(x_i)
    \end{equation}
    is the effective potential barrier height.
    \newline

    \noindent\textbf{Discussion.}  
    This derivation is formally valid only in the small-noise regime.  
    Here, however, the effective noise $\delta_{\text{eff}}$ is rescaled by $(\kappa/\kappa_c)^2$.  
    Whether the Kramers formula remains valid depends on the balance between $\delta$ and $\kappa/\kappa_c$.  

    For example, suppose initially $2\delta \sim 10^{-2}\Delta E_{\text{eff}}$.  
    In the normal case $\kappa\ll \kappa_c$, the escape rate is extremely small, $\Gamma \sim 3\times 10^{-44}$,  
    implying that the system remains in its initial state for all practical purposes: the expected transition time 
    is so large that one would need to wait astronomical timescales to observe a single escape. 
    Now, if $(\kappa/\kappa_c)^2 = 50$, then $2\delta_{\text{eff}} \sim \Delta E_{\text{eff}}/2$,  
    so that $\delta_{\text{eff}} \sim \Delta E_{\text{eff}}/4$.  
    In this regime, the exponential suppression is much weaker, and we obtain $\Gamma \sim 0.13$.
    \newline  

    Thus, the system transitions from near-perfect stability to escaping on timescales of order $10$ in dimensionless units.  
    Even though non-normality rescales the noise amplitude, the dynamics remain within the validity domain of the Kramers problem,  
    while the escape rate can increase by many orders of magnitude.

\subsection{Generalization with Momentum}

    So far, our discussion of non-normal amplification has focused on overdamped Langevin dynamics.  
    To demonstrate that the effect is not restricted to this limit,
    we now consider the more general \emph{underdamped} case,
    where inertia plays a role.  
    Specifically, we study the coupled dynamics
    \begin{subequations}    \label{eq:apx_dyn_under}
    \begin{align}
    \ddot{x} + \gamma\dot{x} &= -\partial_x\phi_x(x) + \kappa\,\partial_y\psi_y(y) + \sqrt{2\delta}\,\eta_x, \\
    \ddot{y} + \gamma\dot{y} &= -\partial_y\phi_y(y) + \kappa^{-1}\partial_x\psi_x(x) + \sqrt{2\delta}\,\eta_y,
    \end{align}
    \end{subequations}
    where $\gamma$ is the linear friction coefficient.  
    This system is the natural generalization of the overdamped equations (Sec.~\ref{apx:effective_potential}),
    now including second-order derivatives.

    \noindent\textbf{Action functional.}
    Following the Onsager-Machlup formulation, the action functional is
    \begin{subequations}
    \begin{align}
    S_\tau[x,y] &= S_\tau^x[x,y] + S_\tau^y[x,y], \\
    S_\tau^x[x,y] &= \tfrac{1}{4}\int_0^\tau \!\left[\ddot{x} + \gamma\dot{x} + \partial_x\phi_x(x) - \kappa\,\partial_y\psi_y(y)\right]^2 d\tau, \\
    S_\tau^y[x,y] &= \tfrac{1}{4}\int_0^\tau \!\left[\ddot{y} + \gamma\dot{y} + \partial_y\phi_y(y) - \kappa^{-1}\partial_x\psi_x(x)\right]^2 d\tau .
    \end{align}
    \end{subequations}
    As in the overdamped case, we expand around a reference point $y=y^*$ satisfying
    \(
        \partial_y\phi_y(y^*)=\partial_y\psi_y(y^*)=0
    \),
    and parametrize $y = y^* + \kappa^{-1}z + \mathcal{O}(\kappa^{-2})$.

    \noindent\textbf{Expansion.}
    To leading order in $\kappa^{-1}$,
    \begin{subequations}
    \begin{align}
    S_\tau^x[x,y] &= \tfrac{1}{4}\int_0^\tau \!\left[\ddot{x} + \gamma\dot{x} + \partial_x\phi_x(x) - \beta z\right]^2 d\tau + \mathcal{O}(\kappa^{-1}),
    \qquad \beta := \partial_y^2\psi_y(y^*), \\[4pt]
    S_\tau^y[x,y] &= \tfrac{1}{4\kappa^2}\int_0^\tau \!\left[\ddot{z} + \gamma\dot{z} + \omega z - \partial_x\psi_x(x)\right]^2 d\tau + \mathcal{O}(\kappa^{-3}),
    \qquad \omega := \partial_y^2\phi_y(y^*).
    \end{align}
    \end{subequations}
    Eliminating $z$ at leading order gives
    \begin{equation}
        z = \frac{1}{\beta}\left[\ddot{x} + \gamma\dot{x} + \partial_x\phi_x(x)\right],
    \end{equation}
    so that
    \begin{subequations}
    \begin{align}
    S_\tau[x,y] &= \frac{1}{\kappa^2}S_\tau[x,z] + \mathcal{O}(\kappa^{-3}), \\
    S_\tau[x,z] &= \tfrac{1}{4}\int_0^\tau \!\left[\ddot{z} + \gamma\dot{z} + \omega z - \partial_x\psi_x(x)\right]^2 d\tau.
    \end{align}
    \end{subequations}
    Inserting the constraint for $z$ yields a fourth-order functional in $x$, consistent with the fact that boundary data must be specified for both positions and velocities $(x,\dot{x},y,\dot{y})$.

    \noindent\textbf{Fast mean-reversion limit.}
    Under the assumption of fast relaxation in $y$ (i.e.~$\omega\gg 1$),  
    and defining the critical ratio
    \(
    \kappa_c := \omega/\beta
    \),
    the action reduces to
    \begin{equation}    \label{eq:apx_under_actiona}
    S_\tau[x,y] \approx \frac{1}{4}\left(\frac{\kappa_c}{\kappa}\right)^2
    \int_0^\tau \!\left[\ddot{x} + \gamma\dot{x} + \partial_x\phi_x(x) - \tfrac{1}{\kappa_c}\partial_x\psi_x(x)\right]^2 d\tau.
    \end{equation}
    This is precisely the underdamped Kramers action, but with a renormalized noise amplitude rescaled by $(\kappa/\kappa_c)^2$ and a modified potential term.

    \noindent\textbf{Equivalent Langevin dynamics.}
    In the same limit, $y$ relaxes to
    \begin{equation}
        y \approx \frac{1}{\beta}\left[\kappa^{-1}\partial_x\psi_x(x) + \sqrt{2\delta}\,\eta_y\right].
    \end{equation}
    Substituting into~\eqref{eq:apx_dyn_under} gives
    \begin{equation}
    \ddot{x} + \gamma\dot{x} \approx -\partial_x\phi_x(x) + \frac{1}{\kappa_c}\partial_x\psi_x(x)
    + \sqrt{2\delta\Bigl(1+(\tfrac{\kappa}{\kappa_c})^2\Bigr)}\,\eta,
    \end{equation}
    which reproduces the effective action~\eqref{eq:apx_under_actiona}.

    \noindent\textbf{Conclusion.}
    Thus, even in the presence of inertia, non-normal coupling rescales the effective noise amplitude in exactly the same way as in the overdamped limit (Sec.~\ref{apx:effective_potential}).  
    The amplification mechanism is therefore a general feature of non-normal stochastic systems, independent of whether momentum is included.

    \subsection{Conclusion}

    In this section, we have analyzed how non-normality amplifies stochastic noise and reshapes the escape dynamics of nonlinear systems.  
    Starting from the action functional and its Hamilton--Jacobi formulation,
    we demonstrated that, in the highly non-normal regime $\kappa \gg \kappa_c$,  
    the quasi-potential $S$ is rescaled by a factor $\kappa^{-2}$.  
    Equivalently, the system experiences an effective noise level $\delta_{\text{eff}} \sim \kappa^2 \delta$,  
    so that non-normality acts as a multiplicative noise-amplification mechanism.  

    We confirmed this result using three complementary approaches:  
    (i) direct minimization of the action functional,  
    (ii) analysis of the leading-order dynamics under a fast-recovery approximation,  
    and (iii) expansion of the Hamilton--Jacobi equation.
    All methods consistently yield the same scaling law,  
    validating both the internal consistency of the analysis and the robustness of the conclusion.  
    Furthermore, by performing a double expansion in $\delta$ and $1/\kappa$,  
    we justified the use of the WKB approximation and clarified that the rescaling applies only at leading order,  
    ensuring asymptotic validity.  

    Finally, we applied these results to the Kramers escape problem,
    showing that the escape rate can increase by many orders of magnitude when $\kappa$ grows,  
    while still remaining within the validity domain of the small-noise approximation.  
    This implies that non-normality not only destabilizes equilibria through deterministic shear,  
    but also drastically enhances stochastic transitions by renormalizing the effective noise scale.  
    As a consequence, systems that would otherwise appear nearly stable may exhibit frequent noise-induced escapes once non-normal amplification is taken into account.  

    In summary, non-normality provides a universal and quantitatively precise mechanism for stochastic amplification:  
    it rescales the quasi-potential barrier by $(\kappa_c/\kappa)^2$ and the noise intensity by $\kappa^2$,  
    leading to exponentially enhanced transition rates in the highly non-normal regime.

\section{Numerical Application}

    To test the validity of our theoretical derivations, we now turn to numerical experiments.  
    We construct a minimal nonlinear model with two potential wells along the reaction ($x$),  
    for which we can explicitly control the degree of non-normality $\kappa$ and compare simulations with theoretical predictions.

    \subsection{Symmetric Potential Well}
    \label{sec:apx_numerical}

    The minimal scalar potential with two stable equilibria and one unstable equilibrium is defined by
    \begin{equation}
        \phi(x) = \frac{\omega}{8}x^2(x^2 - 2),
        \quad \omega > 0 ,
    \end{equation}
    where $\omega$ controls the mean-reversion rate at the stable points.  
    This potential has symmetric wells with stable equilibria at $x=\pm 1$ and an unstable point at $x=0$.  

    For simplicity, we take
    \begin{equation}
        \phi_x(x) = \phi(x), 
        \qquad 
        \phi_y(y) = \phi(y),
    \end{equation}
    so that, in the absence of a solenoidal component,
    the system admits four stable equilibria $(\pm 1,\pm 1)$ and two unstable axes along $x=0$ and $y=0$.

    To introduce a solenoidal contribution that preserves the equilibrium structure, we define
    \begin{equation}
        \psi(x) = \frac{\beta}{6}x(3 - x^2),
    \end{equation}
    for which $\partial_x \psi(x=\pm 1) = 0$.  
    Hence, the solenoidal component does not alter equilibrium stability.  
    We then set $\psi_x(x) = \psi(x)$ and $\psi_y(y) = \psi(y)$.  

    The full system reads
    \begin{subequations}
        \begin{align}
            \dot{x} &= -\partial_x \phi(x) + \kappa \partial_y \psi(y) + \sqrt{2\delta}\,\eta_x , \\
            \dot{y} &= -\partial_y \phi(y) + \kappa^{-1}\partial_x \psi(x) + \sqrt{2\delta}\,\eta_y ,
        \end{align}
    \end{subequations}
    where $\eta_x,\eta_y$ are independent unit white noises.
    
    One can also note that, in the neighborhood of any stable equilibrium, the Jacobian of the generalized force field takes the form
    \begin{equation}
        \J =
        \begin{pmatrix}
            -\omega & \pm \beta \\
            \pm \beta & -\omega
        \end{pmatrix},
    \end{equation}
    where the sign $\pm$ depends on which equilibrium $(x,y)=(\pm 1,\pm 1)$ the system is linearized around.  
    The corresponding eigenvalues can be written, without loss of generality, as
    \begin{equation}
        \lambda_\pm = -\omega \pm \beta\chi,
    \end{equation}
    with $\chi=1$ for the equilibria on the diagonal ($x=y=\pm 1$)  
    and $\chi=i$ for the off-diagonal equilibria ($x=-y=\pm 1$).  

    Recall that we defined the critical degree of non-normality as $\kappa_c = \omega/\beta$.  
    This recovers the same notion of criticality introduced in \cite{troude2025}:  
    pseudo-critical amplification occurs when $\kappa>\kappa_c$,  
    with $\kappa_c$ scaling proportionally to the distance from criticality ($\omega$)  
    and inversely to the degree of degeneracy ($\beta$).  
    Thus, the framework of unified amplification in linear systems naturally extends to the nonlinear setting considered here.  

    Finally, to ensure the stability of the diagonal equilibria ($x=y=\pm 1$),  
    the parameters must satisfy $\omega > |\beta| \geq 0$.
    \newline

    In the regime where $y \approx \pm 1$ is ``almost stable,''  
    the effective dynamics of $x$ reduces to 
        \begin{equation} \label{eq:dyn_eff_apx}
        \dot{x} = -U_{\text{eff}}'(x) + \sqrt{2\delta_{\text{eff}}}\,\eta ,
    \end{equation}
    with
    \begin{equation}
        U_{\text{eff}}(x) = \phi(x) \mp \frac{1}{\kappa_c}\psi(x),
        \qquad
        \delta_{\text{eff}} = \delta\left[1 + \left(\frac{\kappa}{\kappa_c}\right)^2\right],
        \qquad
        \kappa_c = \frac{\omega}{\beta}.
    \end{equation}
    The sign in $U_{\text{eff}}(x)$ depends on the choice $y^* = \pm 1$.  

    For an escape from $x_i=\pm 1$ to $x_f=0$,  
    the effective barrier and prefactor are
    \begin{equation} \label{eq:eff_sym_apx}
        \Delta E_{\text{eff}} = \omega\left[\frac{1}{8} \mp \frac{1}{2\kappa_c^2}\right],
        \qquad
        C = \frac{1}{2\pi}\sqrt{\frac{\omega}{2}\left(1 \mp \frac{1}{\kappa_c^2}\right)} .
    \end{equation}
    Thus, the dynamics is controlled by four parameters:
    \begin{itemize}
        \item $\omega$ : mean-reversion rate around equilibria.
        \item $\delta$ : amplitude of the input noise.
        \item $\kappa_c$ : critical threshold for the restoring amplitude to non-normal shear.
        \item $\kappa$ : actual strength of the non-normal shear.
    \end{itemize}

    \subsection{Numerical Result}

\begin{figure}
        \centering
        \includegraphics[width=\textwidth]{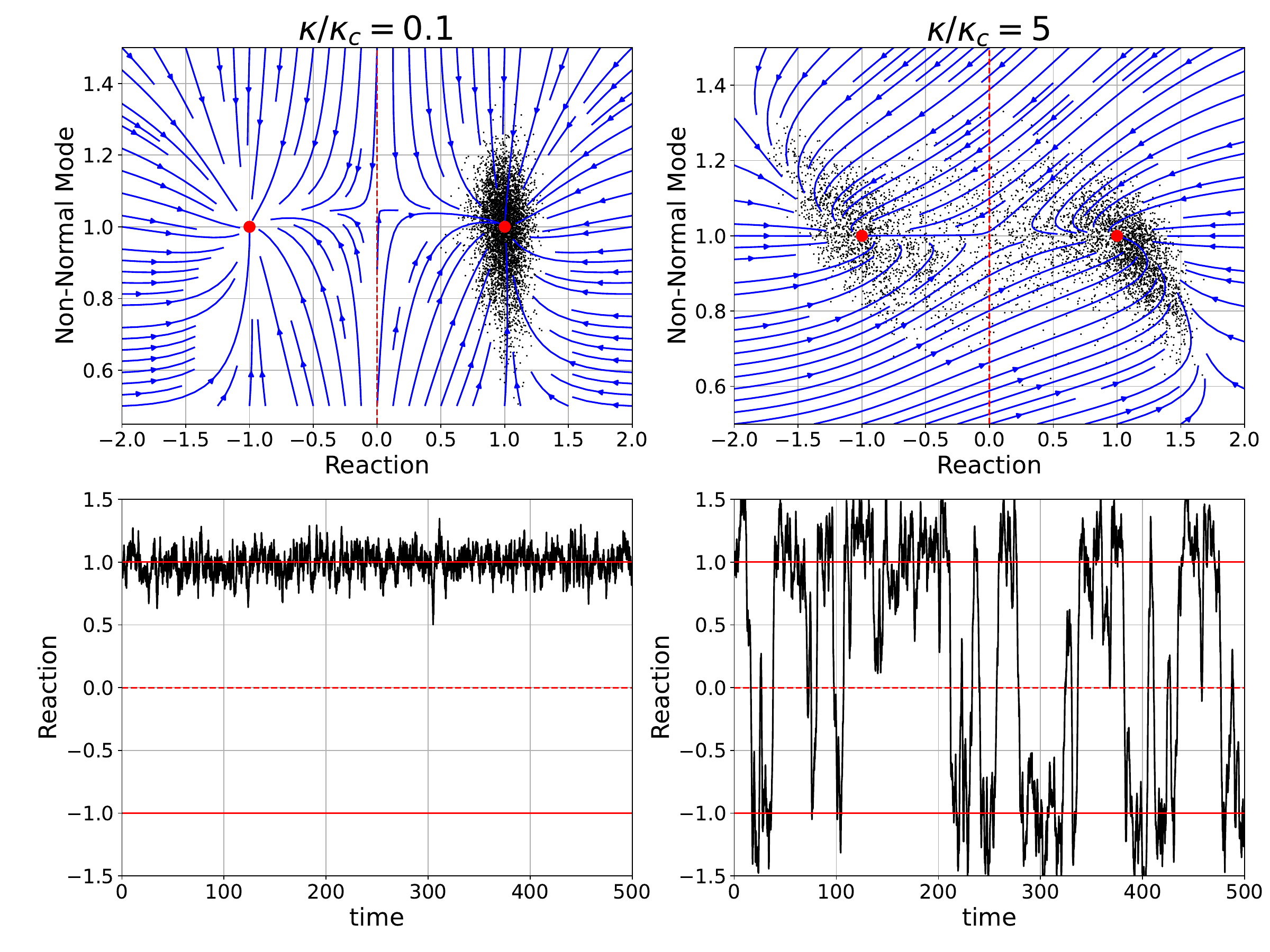}
        \caption{
            Simulation of a nonlinear two-dimensional system described in Section~\ref{sec:apx_numerical}, 
            with parameters $\omega=1$, $\delta=0.01$, and $\kappa_c=10$. 
            The left panels correspond to $\kappa=\kappa_c/10$, while the right panels correspond to $\kappa=5\kappa_c$. \\
            Top panels: dynamics in phase space, where the horizontal axis denotes the reaction coordinate ($x$) 
            and the vertical axis the non-normal mode ($y$). 
            Red dots mark the stable equilibria $(x,y)=(\pm 1,1)$, 
            blue arrows indicate the force vector field, 
            and the dashed red line at $x=0$ denotes the unstable manifold along the reaction direction. \\
            Bottom panels: time series of the reaction variable ($x$). 
            Continuous red lines mark the stable equilibria at $x=\pm 1$, 
            and the dashed red line marks the unstable equilibrium at $x=0$. \\
            All simulations are performed over a time horizon $T=500$ 
            with integration step $\Delta t=0.1$, 
            corresponding to $N=5000$ time steps.
        }
        \label{fig:dynamic_apx}
    \end{figure}

    Figure~\ref{fig:dynamic_apx} shows trajectories for $\delta = 10^{-2}$, $\omega=1$, $\kappa_c=10$,  
    with $\kappa=\kappa_c$ (left) and $\kappa=5\kappa_c$ (right).  
    The top row displays the phase space trajectories, while the bottom row shows the dynamics of $x$.  
    For $\kappa=\kappa_c$, the system remains trapped in a single well for the entire simulation,  
    whereas for $\kappa=5\kappa_c$, it transitions $\sim 20$ times between $x=\pm 1$ over the duration 
    $T=500$ of the simulations.  

\begin{figure}
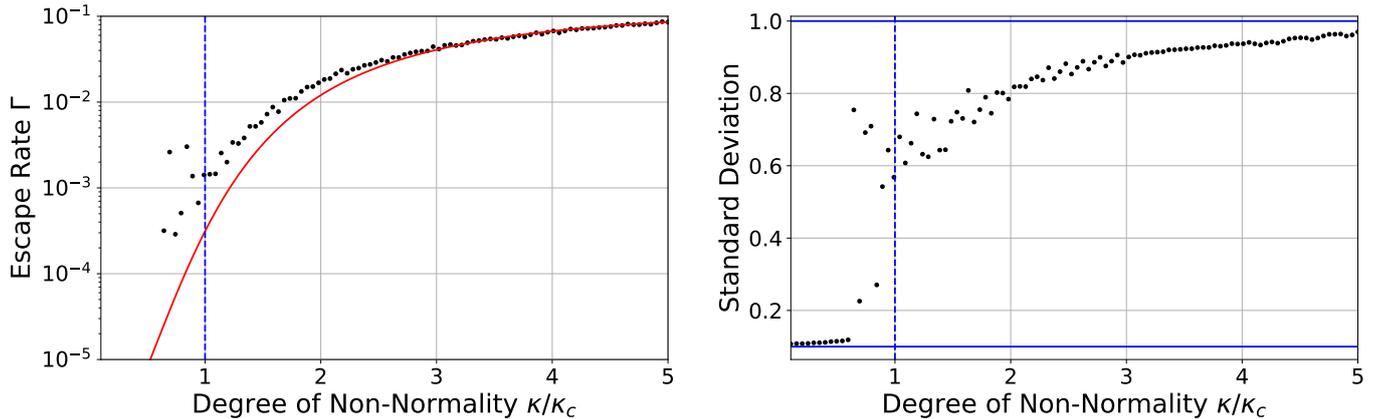

        \centering
        \begin{subfigure}{0.49\textwidth}
            \centering
            \includegraphics[width=\textwidth]{escape_rate.pdf}
        \end{subfigure}
        \begin{subfigure}{0.49\textwidth}
            \centering
            \includegraphics[width=\textwidth]{std.pdf}
        \end{subfigure}
        \caption{
            Escape rate (left panel) and standard deviation (right panel) of the reaction variable ($x$) 
            for the dynamics described in Section~\ref{sec:apx_numerical}, 
            with parameters $\omega=1$, $\delta=0.01$, and $\kappa_c=10$, 
            as a function of $\kappa$. 
            Simulations are performed for $\kappa$ ranging from $\kappa_c/10$ to $5\kappa_c$. \\
            Each simulation runs over a total time $T=10^{5}$ 
            with integration step $\Delta t=0.1$, 
            corresponding to $N=10^{6}$ data points. 
            The escape rate is estimated as $\Gamma = 1/\langle\tau\rangle$, 
            where $\langle\tau\rangle$ is the mean first-passage time from $x>0$ to $x<0$ (or vice versa). \\
            Black dots denote numerical measurements, 
            and the red curve shows the theoretical escape rate given by Eq.~\eqref{eq:escape_apx}. 
            The vertical dashed blue line marks the critical value $\kappa=\kappa_c$. 
            In the right panel, the lower horizontal blue line indicates the theoretical standard deviation 
            of an Ornstein-Uhlenbeck approximation near equilibrium, $\sqrt{\delta/\omega}=0.1$, 
            while the upper blue line ($=1$) corresponds to the variance of a process equally likely to be near $x=\pm 1$.
        }
        \label{fig:escape_apx}
    \end{figure}

    A systematic scan over $\kappa$, shown in Figure~\ref{fig:escape_apx},  
    confirms this transition.  
    For $\kappa<\kappa_c$, escapes are very rare as the transition rate is exceedingly small,  
    and the variance of $x$ remains close to the Ornstein-Uhlenbeck prediction $\sqrt{\delta/\omega} = 1/10$.  
    At $\kappa=\kappa_c$, a qualitative shift occurs:  
    the system begins to transition between wells while the input noise level remains constant at a very small level,  
    and the variance abruptly increases to $\sim 1$.  
    For $\kappa>\kappa_c$, the measured escape rate converges to the theoretical prediction \eqref{eq:eff_sym_apx},  
    validating both the rescaling $\delta_{\text{eff}} \sim \kappa^2\delta$ and the overall large-deviation framework.

    \subsection{Asymmetric Potential Well}

    In the previous section, we studied the symmetric case, 
    where the dynamics of the reaction variable $x$ is nearly symmetric, 
    so that the transition rates between the two stable equilibria are identical.  
    We now introduce an asymmetry by modifying the scalar and solenoidal potentials along $x$, defined as
    \begin{equation}
        \phi_x(x) = \frac{\omega}{1+\Delta}\,\frac{1}{4}x^2\left[x^2 + \frac{4}{3}(\Delta - 1)x - 2\Delta\right],
        \qquad
        \psi_x(x) = \frac{\beta}{1+\Delta}\,\frac{1}{3}x\left[x^2 + \frac{3}{2}(\Delta -1)x - 3\Delta\right],
    \end{equation}
    whose derivatives are
    \begin{equation}
        \partial_x\phi_x(x) = \frac{\omega}{1+\Delta}\,x(x-1)(x+\Delta),
        \qquad
        \partial_x\psi_x(x) = \frac{\beta}{1+\Delta}(x-1)(x+\Delta).
    \end{equation}

    For $\Delta>1$, the system retains one unstable equilibrium at $x=0$, 
    but the two stable equilibria are now located at $x=1$ and $x=-\Delta$.  
    Crucially, the heights of the potential barriers separating these equilibria are no longer identical.  

    Keeping the potentials along the $y$-direction unchanged from Section~\ref{sec:apx_numerical}, 
    the critical degree of non-normality remains $\kappa_c=\omega/\beta$.  
    The effective potential barrier from $x=1$ to $x=0$ is then
    \begin{equation}
        \Delta E_{\text{eff},1} = \frac{\omega}{1+\Delta}
        \left[
            \frac{1}{12}(1+2\Delta) - \frac{1}{6\kappa_c^2}(1+3\Delta)
        \right].
    \end{equation}
    For $\Delta=1$, this reduces to the symmetric case of Eq.~\eqref{eq:eff_sym_apx}, 
    while in the limit $\Delta\to\infty$, the barrier remains of order $\mathcal{O}(1)$.  

    In contrast, the effective barrier from $x=-\Delta$ to $x=0$ is
    \begin{equation}
        \Delta E_{\text{eff},-\Delta} = \frac{\omega\Delta^2}{1+\Delta}
        \left[
            \frac{1}{12}(\Delta+2) + \frac{1}{6\kappa_c^2}(\Delta+3)
        \right],
    \end{equation}
    which grows as $\mathcal{O}(\Delta^2)$ in the limit $\Delta\to\infty$.  

    Thus, the asymmetric potential well provides a bistable system in which non-normality can induce sufficiently large fluctuations to overcome the barrier in one direction,
    while the reverse transition remains exponentially suppressed.  
    This construction highlights how non-normal amplification can break reversibility in noise-induced transitions.
    
    \subsection{Conclusion}

    Through the numerical study of symmetric and asymmetric bistable potentials, 
    we have validated the theoretical prediction that the quasi-potential is rescaled by a factor $(\kappa_c/\kappa)^2$, 
    where the critical degree of non-normality $\kappa_c=\omega/\beta$ 
    acts as the threshold separating regimes where non-normal amplification is negligible ($\kappa<\kappa_c$) 
    or dominant ($\kappa>\kappa_c$).  
    This provides a clear interpretation of $\kappa_c$:
    it links the strength of non-normal shear to the intrinsic stability of the system, 
    so that criticality in the reversible (gradient) part of the dynamics and the non-normal amplification mechanism are tied together.  

    Studying bistable potentials is particularly insightful, 
    since transitions between stable equilibria are the most direct manifestation of stochastic fluctuations.  
    In the symmetric case, we observed that the onset of non-normality induces a sudden increase in transition rates, 
    while in the asymmetric case, the same mechanism selectively amplifies fluctuations in one direction, 
    leading to irreversible dynamics where escape is favored only from one potential well.  
    This demonstrates that non-normal amplification not only accelerates noise-induced transitions, 
    but can also fundamentally alter the symmetry and reversibility of the system's long-term behavior.  

    Finally, although we restricted the analysis to quartic potentials generating bistability, 
    the framework is not limited to this case.  
    Higher-order polynomials can be used to construct systems with multiple stable equilibria, 
    thus extending the analysis to multistable landscapes.  
    In all cases, however, the transitions along the reaction coordinate are expected to obey the same rule: 
    their rates are governed by the balance between the potential restoring force and the shear strength
    (measured respectively by $\omega$ and $\beta$) combined in the critical degree of non-normality $\kappa_c=\omega/\beta$
    and non-normal shear (measured by $\kappa$).
    This shows that $\kappa/\kappa_c$ provides a universal coefficient controlling the strength and impact of non-normal amplification 
    across a broad class of nonlinear stochastic systems.

\section{Application to DNA Methylation}

    DNA methylation is a key epigenetic mechanism that modulates gene regulation and cellular identity. 
    Yet methylation patterns can switch states on unexpectedly fast timescales
    -- sometimes within minutes in response to environmental cues --
    challenging predictions from classical variational Kramers-type models \cite{busto2020stochastic}.
    Several biological features plausibly contribute to this rapidity:
    (i) chromatin architecture locally boosts DNMT access and activity,
    creating methylation hotspots \cite{baubec2015genomic},
    (ii) stochastic metabolic fluctuations (e.g., transient surges in S-adenosylmethionine)
    amplify reaction propensities \cite{busto2020stochastic},
    and (iii) positive feedback, whereby methylation at one CpG promotes methylation in neighboring regions,
    supports rapid propagation of marks (chemical modifications on DNA or histones that modulate gene expression without changing sequence) \cite{day2010dna,kim2017dna}.
    External signals (oxidative stress, pathway activation) further reshape the methylation landscape dynamically.

    Viewed through the lens of \emph{non-normal} stochastic dynamics, these observations admit a parsimonious explanation. 
    In non-normal systems, the solenoidal/rotational component of the force field transiently amplifies perturbations, which 
    \emph{renormalizes the effective noise level} and, in turn, the escape kinetics. 
    Concretely, in the small-noise limit used throughout this SM,
    the dynamics along the reaction coordinate obeys an effective 1D Langevin equation,
    so that the Kramers rate inherits the standard form with $\delta$ replaced by $\delta_{\text{eff}}$
    i.e. see \eqref{eq:deltaeff} and \eqref{eq:escape_apx}. 
    This renormalization explains how methylation transitions can occur on minute timescales despite modest thermal noise:
    transient amplification effectively raises the temperature experienced along the escape path,
    while preserving the system's bistability structure.

    Empirical features of DNA methylation are consistent with the three hallmarks of non-normality:
    \begin{enumerate}[(i)]
        \item \textbf{Asymmetry.} DNMTs preferentially target specific sequence and chromatin contexts, biasing local dynamics.
        \item \textbf{Hierarchy.} Local positive feedback enables cascading spread from hemimethylated to fully methylated regions.
        \item \textbf{Stochastic fluctuations.} Variability in methyl-donor availability and enzymatic activity injects extrinsic and intrinsic noise.
    \end{enumerate}
    Together, these place methylation dynamics squarely within the class of non-variational, highly non-normal systems. 
    In what follows, we make this connection explicit by embedding an established bistable model of CpG dyads within our framework,
    adding Langevin noise, and quantifying how non-normal amplification ($\kappa/\kappa_c$) accelerates transitions while maintaining bistability.

\subsection{Model}

    To make the connection with DNA methylation explicit,
    we start from the nonlinear model in \cite{zagkos2019}. 
    This framework was developed to describe the coexistence of unmethylated, hemimethylated, and methylated CpG dyads,
    thereby rationalizing the experimentally observed bistability of methylation. 
    This model captures how localized interactions and cooperative enzymatic processes drive transitions between hypo- and hypermethylated states. 
    In its original form, the dynamics are deterministic, and noise is absent;
    however, the authors noted the importance of incorporating stochasticity to reflect uncertainty in methylation levels. 
    Here, by uncertainty we mean fluctuations around the attractors corresponding to unmethylated, hemimethylated, and fully methylated CpG dyads. 
    We extend their formulation by explicitly embedding stochastic fluctuations within an overdamped Langevin framework,
    which allows us to characterize not only the variance of methylation levels but also the noise-driven transition rates between epigenetic states.

    \noindent\textbf{Stochastic embedding.}
    The model in \cite{zagkos2019} can be naturally embedded into our non-normal stochastic framework. 
    Introducing Gaussian noise terms, we obtain the coupled equations
    \begin{equation}    \label{eq:dnmt_dynamic}
    \begin{cases}
    \dot{x}_1 = -a_{13}x_1 ^3 + a_{12} C x_1^2 - a_{11} x_1 + a_{10}C - b_{13}x_1 ^2 x_3 - b_{11}x_3 + \sqrt{2\delta}\,\eta_1, \\
    \dot{x}_3 = -a_{33}x_3 ^3 + a_{32}C x_3^2 - a_{31} x_3 + a_{30}C - b_{33}x_3 ^2 x_1 - b_{31}x_1 + \sqrt{2\delta}\,\eta_3,
    \end{cases}
    \quad\text{with}\quad
    \eta_1,\eta_3\overset{iid}{\sim}\mathcal{N}(0,1),
    \end{equation}
    where $x_1$ and $x_3$ denote the numbers of unmethylated and methylated dyads, respectively, and the hemimethylated count is 
    \(
    x_2 = C - x_1 - x_3
    \),
    with $C>0$ the total number of CpG dyads.

    \noindent\textbf{Hyper-parameters.}
    The coefficients $a_{ij}$ and $b_{ij}$ are defined in terms of transition rates $k_{ij}$ as
    \begin{subequations}    \label{eq:hyper_para_apx}
    \begin{align}
    &a_{13} = k_{32}-k_{12}, \quad
    a_{12} = k_{32}, \quad
    a_{11} = k_{11} + k_{31} + \tfrac{1}{2}D, \quad
    a_{10} = k_{31} + \tfrac{1}{2}D, \quad
    b_{13} = k_{32}, \quad
    b_{11} = k_{31} + \tfrac{1}{2}D, \\
    &a_{33} = k_{22} - k_{42}, \quad
    a_{32} = k_{22}, \quad
    a_{31} = k_{21} + k_{41} + D, \quad
    a_{30} = k_{21}, \quad
    b_{33} = k_{22}, \quad
    b_{31} = k_{21},
    \end{align}
    \end{subequations}
    with $k_{ij}$ denoting reaction rates, in the notation of \cite{zagkos2019}. 
    Biologically, $k_{21}$ represents the effective methylation rate of hemimethylated dyads,
    $k_{31}$ the active demethylation rate, and $D$ the cell-division rate (which contributes to passive demethylation when maintenance is incomplete). 
    The parameter values used in \cite{zagkos2019} are summarized in Table~\ref{tab:parameters_apx}.

    \begin{table}[h!]
    \centering
    \begin{tabular}{|c|c|c|c|c|c|c|c|c|c|}
    \hline
    $k_{11}$ & $k_{12}$ &  $k_{21}$ &  $k_{22}$ & $k_{31}$ & $k_{32}$ & $k_{41}$ & $k_{42}$ & $C$ & $D$ \\
    \hline
    $2.1$ & $2 \times 10^{-5}$ & $10$ & $10^{-2}$ & $1$ & $10^{-2}$ & $4$ & $2 \times 10^{-4}$ & $100$ & $1$ \\
    \hline
    \end{tabular}
    \caption{Parameter values used in the model of \cite{zagkos2019}.}
    \label{tab:parameters_apx}
    \end{table}

    \noindent\textbf{Simplifications.}
    Using Table~\ref{tab:parameters_apx} together with definitions \eqref{eq:hyper_para_apx},
    several simplifications follow
    \begin{equation}
    C \gg 1, 
    \qquad
    a_{13}\approx a_{33} \approx a_{12} = a_{32} = b_{13} = b_{33}, 
    \qquad
    a_{11}\approx a_{10}\approx b_{11},
    \qquad
    a_{31}\approx a_{30} \approx b_{31}.
    \end{equation}
    Since $x_1,x_3 \in (0,C)$, it is natural to rescale
    \begin{equation}
    x_1 = C y_1, 
    \qquad
    x_3 = C y_3,
    \end{equation}
    yielding
    \begin{equation}
    \begin{cases}
    \dot{y}_1 = - C^2 y_1 ^2\!\left[a_{13} y_1 + b_{13} y_3 - a_{12}\right] - a_{11} y_1 + a_{10} - b_{11}y_3 + C^{-1}\sqrt{2\delta}\,\eta_1, \\
    \dot{y}_3 = - C^2 y_3 ^2\!\left[a_{33}y_3 + b_{33}y_1 - a_{32}\right] - a_{31} y_3 + a_{30} - b_{31}y_1 + C^{-1}\sqrt{2\delta}\,\eta_3 .
    \end{cases}
    \end{equation}

    From the parameter hierarchy, note that $a_{13},a_{33}\sim C^{-1}$,
    and $k_{12}$, $k_{32}$ and $k_{42}$ are orders of magnitude lower than the other parameters.
    This leads to the reduced form
    \begin{equation}    \label{eq:model_dna_reduced}
    \begin{cases}
    \dot{y}_1 = - \bigl[C y_1 ^2 + \omega_1\bigr]\bigl(y_1 + y_3 - 1\bigr) - k y_1 + C^{-1}\sqrt{2\delta}\,\eta_1, \\
    \dot{y}_3 = - \bigl[C y_3 ^2 + \omega_3\bigr]\bigl(y_1 + y_3 - 1\bigr) - k y_3 + C^{-1}\sqrt{2\delta}\,\eta_3,
    \end{cases}
    \end{equation}
    where
    \begin{equation}
    \omega_i := a_{i0} = b_{i1}, 
    \qquad
    k \approx k_{11},k_{41}.
    \label{rhybgrqv87}
    \end{equation}
    Here, we consolidate $k_{11}$ and $k_{41}$ into a single effective parameter $k$, reducing the dimensionality of the hyper-parameter set.

    \noindent\textbf{Time rescaling.}
    Introducing a rescaling $t \mapsto Ct$ simplifies the system to
    \begin{equation}     \label{eq:model_dna_final}
    \begin{cases}
    \dot{y}_1 = - y_1 ^2 \bigl(y_1 + y_3 - 1\bigr) - C^{-1}\!\left[\omega_1\bigl(y_1 + y_3 - 1\bigr) + k y_1\right] + C^{-3/2}\sqrt{2\delta}\,\eta_1, \\
    \dot{y}_3 = - y_3 ^2 \bigl(y_1 + y_3 - 1\bigr) - C^{-1}\!\left[\omega_3\bigl(y_1 + y_3 - 1\bigr) + k y_3\right] + C^{-3/2}\sqrt{2\delta}\,\eta_3 .
    \end{cases}
    \end{equation}

    \noindent\textbf{Summary.}
    Equation~\eqref{eq:model_dna_final} is the reduced stochastic model that we analyze throughout this section. 
    In the next steps, we (i) identify stable and unstable equilibria,
    (ii) characterize the non-normal mode and its reaction variable,
    and (iii) quantify the strength of the non-normal shear that determines when non-normal amplification of transition rates occurs.

\subsection{Equilibrium Points}

    Our next step is to approximate the stable equilibria of the reduced model~\eqref{eq:model_dna_final}. 
    Identifying these equilibria is crucial for two reasons:
    (i) the line connecting the two stable fixed points defines the reaction coordinate,
    and (ii) the non-normal mode must be orthogonal to this line. 
    This construction provides a geometric way to separate the reaction direction from the non-normal shear.

    \noindent\textbf{Location of equilibria.}
    From \cite{zagkos2019}, the system admits two stable fixed points: one with $x_1 \sim C$, $x_3 \sim 1$, and the other with $x_1 \sim 1$, $x_3 \sim C$. 
    In the rescaled variables $y_i = x_i / C$, and in the large-$C$ limit, these equilibria correspond to
    \(
    (y_1,y_3) \approx (1,C^{-1})
    \)
    and
    \(
    (y_1,y_3) \approx (C^{-1},1)
    \).
    Both are close to the line $y_1+y_3=1$. 
    We therefore introduce rotated coordinates $(z_1,z_3)$ defined by
    \begin{equation}    \label{eq:apx_z_to_y}
    \begin{cases}
    z_1 = \tfrac{1}{\sqrt{2}}\,(y_1+y_3-1), \\[4pt]
    z_3 = \tfrac{1}{\sqrt{2}}\,(y_1-y_3),
    \end{cases}
    \quad \Longleftrightarrow \quad
    \begin{cases}
    y_1 = \tfrac{1}{\sqrt{2}}\,(z_1+z_3) + \frac{1}{2}, \\[4pt]
    y_3 = \tfrac{1}{\sqrt{2}}\,(z_1-z_3) + \frac{1}{2}.
    \end{cases}
    \end{equation}
    This transformation consists of a translation plus a rotation,
    so it preserves the system's non-normality.

    \noindent\textbf{Dynamics in the rotated basis.}
    In the $(z_1,z_3)$ variables, the deterministic part of the dynamics becomes
    \begin{equation}    \label{eq:apx_dyn_z}
    \begin{cases}
    \dot{z}_1 = -\bigl[\left(z_1+\frac{1}{\sqrt{2}}\right)^2 + z_3^2\bigr]\,z_1 
    - C^{-1}(\omega_+ + k)z_1 - C^{-1}\frac{k}{\sqrt{2}}, \\
    \dot{z}_3 = -2\left(z_1+\frac{1}{\sqrt{2}}\right)z_3 z_1 
    - C^{-1}\bigl(\omega_- z_1 + k z_3\bigr),
    \end{cases}
    \qquad 
    \omega_\pm = \omega_1 \pm \omega_3.
    \end{equation}

    \noindent\textbf{Asymptotic expansion.}
    Because equilibria satisfy $\dot{z}_1=\dot{z}_3=0$, we expand
    \begin{equation}
    z_1 = z_1^{(1)} C^{-1} + z_1^{(2)} C^{-2} + \mathcal{O}(C^{-3}),
    \qquad
    z_3 = z_3^{(0)} + z_3^{(1)} C^{-1} + \mathcal{O}(C^{-2}).
    \end{equation}
    Substituting into~\eqref{eq:apx_dyn_z} and matching powers of $C^{-1}$ gives recursive equations for $z_1^{(j)}$ and $z_3^{(j)}$.

    \noindent\textbf{Order $\mathcal{O}(C^{-1})$.}
    At leading order,
    \begin{equation}
    \begin{cases}
    (\frac{1}{2}+(z_3^{(0)})^2)\,z_1^{(1)} + \frac{k}{\sqrt{2}} = 0, \\
    (\sqrt{2}z_1^{(1)}+k)\,z_3^{(0)} = 0,
    \end{cases}
    \quad \Longrightarrow \quad
    \begin{cases}
    z_1^{(1)} = -\sqrt{2}k, \\ z_3^{(0)}=0,
    \end{cases}
    \quad\text{or}\quad
    \begin{cases}
    z_1^{(1)} = -\tfrac{k}{\sqrt{2}}, \\ z_3^{(0)}=\pm \frac{1}{\sqrt{2}}.
    \end{cases}
    \end{equation}

    \noindent\textbf{Order $\mathcal{O}(C^{-2})$.}
    Proceeding to next order yields
    \begin{equation}
    \begin{cases}
    \left(\frac{1}{2}+(z_3^{(0)})^2\right)z_1^{(2)} + 2z_3^{(0)} z_1^{(1)} z_3^{(1)} 
    = -z_1^{(1)}\left(\sqrt{2}z_1^{(1)}+\omega_+ + k\right), \\
    \sqrt{2}z_3^{(0)} z_1^{(2)} + (\sqrt{2}z_1^{(1)}+k)z_3^{(1)} 
    = -z_1^{(1)}(2z_1^{(1)} z_3^{(0)} + \omega_-).
    \end{cases}
    \end{equation}
    From this, the equilibria are
    \begin{equation}
    \begin{cases}
    z_{1}^{(0)} = -\sqrt{2}k C^{-1} + \mathcal{O}(C^{-2}), \\[4pt]
    z_{3}^{(0)} = -\sqrt{2}\omega_- C^{-1} + \mathcal{O}(C^{-2}),
    \end{cases}
    \qquad \text{or} \qquad
    \begin{cases}
    z_{1}^{(\pm)} = -\tfrac{k}{\sqrt{2}} C^{-1} + \mathcal{O}(C^{-2}), \\[4pt]
    z_{3}^{(\pm)}  = \pm \frac{1}{\sqrt{2}} - \tfrac{1}{\sqrt{2}}C^{-1}(\pm k \pm \omega_+ - \omega_-) + \mathcal{O}(C^{-2}).
    \end{cases}
    \end{equation}

    \noindent\textbf{Jacobian analysis.}
    Linearizing~\eqref{eq:apx_dyn_z} about each equilibrium yields Jacobians
    \begin{subequations}
    \begin{align}
    \J_0 &= -
    \begin{pmatrix}
        1/2 & 0 \\ 0 & 0
    \end{pmatrix}
    + C^{-1}
    \begin{pmatrix}
    3k-\omega_+ & 0 \\
    \omega_- & k
    \end{pmatrix}
    + \mathcal{O}(C^{-2}), \\
    \J_\pm &= -
    \begin{pmatrix}
    1 & 0 \\
    \pm 1 & 0
    \end{pmatrix}
    + C^{-1}
    \begin{pmatrix}
    2k\pm \omega_- & \pm(-k) \\
    \pm(3k+\omega_+) - \omega_- & 0
    \end{pmatrix}
    + \mathcal{O}(C^{-2}),
    \end{align}
    \end{subequations}
    where $\J_0$ and $\J_\pm$ are respectively the Jacobian estimated at $\z_0 = (z_{1,0}\,,\,z_{3,0})$,
    and $\z_\pm = (z_{1,\pm}\,,\,z_{3,\pm})$.

    Their eigenvalues are
    \begin{equation}
    \begin{cases}
    \lambda_{+,0} = -\frac{1}{2}+ C^{-1}(3k-\omega_+) + \mathcal{O}(C^{-2}), \\[4pt]
    \lambda_{-,0} = C^{-1}k + \mathcal{O}(C^{-2}),
    \end{cases}
    \qquad
    \begin{cases}
    \lambda_{+,\pm} = -1 - C^{-1}(-3k \pm \omega_-) + \mathcal{O}(C^{-2}), \\[4pt]
    \lambda_{-,\pm} = -2k C^{-1} + \mathcal{O}(C^{-2}).
    \end{cases}
    \end{equation}

    \noindent\textbf{Conclusion.}
    The point $\z_0$ is unstable, while $\z_\pm$ are stable equilibria. 
    We interpret $\z_+$ as the methylated state and $\z_-$ as the unmethylated state.
    To obtain transition between the two states $\z_\pm$,
    we need to estimate if the system is non-normal,
    and it is required to have the reaction aligned along the $z_3$-axis.

\subsection{Non-Normality of the System}

    To quantify the non-normality of the reduced dynamics~\eqref{eq:model_dna_final}, 
    we compute the eigenvectors of the Jacobian at each equilibrium point $\z_0$ (unstable) and $\z_\pm$ (stable).
    Expanding to order $\mathcal{O}(C^{-1})$ yields
    \begin{subequations}
    \begin{align}
    \p_{+,\pm} &= \tfrac{1}{\sqrt{2}}
    \begin{pmatrix} 1 \\ \pm 1 \end{pmatrix}
    + \frac{1}{\sqrt{2}} C^{-1}
    \begin{pmatrix}
        -3k\pm \left(\omega_- - \frac{\omega_+}{2}\right)  \\ 
        \pm\left(3k+\omega_+\right) - \omega_- - \frac{\omega_+}{2}
     \end{pmatrix}
    + \mathcal{O}(C^{-2}),
    &
    \p_{-,\pm} &=
    \begin{pmatrix} 0 \\ 1 \end{pmatrix}
    \pm 2C^{-1}k
    \begin{pmatrix} 1 \\ 0 \end{pmatrix}
    + \mathcal{O}(C^{-2}), \\[6pt]
    \p_{+,0} &= 
    \begin{pmatrix}
        1 \\ 0
    \end{pmatrix}
    -2C^{-1}\omega_-
    \begin{pmatrix} 0 \\ 1 \end{pmatrix}
    + \mathcal{O}(C^{-2}),
    &
    \p_{-,0} &= \begin{pmatrix} 0 \\ 1 \end{pmatrix}
    +2C^{-1}k
    \begin{pmatrix}
        1 \\ 0
    \end{pmatrix}
     + \mathcal{O}(C^{-2}).
    \end{align}
    \end{subequations}

    \noindent\textbf{Condition number and non-normal index.}
    The degree of non-normality is captured by the condition number of the eigenbasis \cite{troude2024},
    \begin{equation}
    \kappa_i = \sqrt{\frac{1 + |\p_{+,i}\cdot\p_{-,i}|}{1 - |\p_{+,i}\cdot\p_{-,i}|}},
    \end{equation}
    where $i\in\{0,\pm\}$ denotes the equilibrium point~\cite{troude2025}.
    We obtain
    \begin{equation}
    \kappa_0 = 1 + 2C^{-1}\left|k-\omega_+\right| + \mathcal{O}(C^{-2}),
    \qquad
    \kappa_\pm = \left(\sqrt{2} + 1\right)\left[1 - \sqrt{2}C^{-1}\left(\pm\left(5k+\omega_+\right)-\omega_--\frac{\omega_+}{2}\right)\right] +\mathcal{O}(C^{-2}),
    \end{equation}
    therefore, close to the stable equilibrium, the system is always non-normal,
    but near the unstable equilibrium the system is almost normal.
    In all cases $\kappa_i>1$, confirming that the system is non-normal near each equilibrium.

    A convenient scalar measure is the \emph{non-normal index}~\cite{troude2025}
    \begin{equation}
    K_i = \tfrac{1}{2}\left(\kappa_i - \kappa_i^{-1}\right).
    \end{equation}
    Comparing $K_i$ with its critical threshold
    \begin{equation}
    K_{c,i} := \sqrt{\frac{\sqrt{\alpha_i^2-1}}{\alpha_i-\sqrt{\alpha_i^2-1}}},
    \qquad
    \alpha_i = \left|\tfrac{\lambda_{+,i}+\lambda_{-,i}}{\lambda_{+,i}-\lambda_{-,i}}\right|,
    \end{equation}
    identifies whether the system is \emph{pseudo-critical},
    meaning that transient perturbations are amplified along the reaction coordinate before decaying.
    For the stable equilibria, we obtain
    \begin{equation}
    K_{c,\pm} = 2^{3/4}\left(\tfrac{k}{C}\right)^{1/4} + \mathcal{O}(C^{-3/4}),
    \end{equation}
    so that
    \(
        K_\pm/K_{c,\pm} = \mathcal{O}(C^{1/4}).
    \)
    Thus, the non-normal amplification grows with system size $C$.  
    The unstable equilibrium $\z_0$ is naturally unstable,
    so perturbations there grow exponentially regardless of non-normality.

    \noindent\textbf{Conclusion.}
    The DNA methylation model is strongly non-normal near its equilibria.  
    This ensures transient deviations in the linearized dynamics, raising the key question:  
    do these deviations enhance the transition rates between the unmethylated and methylated states?

\subsection{Reaction and Non-Normal Mode}

    The stable equilibria lie near $z_1\approx 0$, $z_3\approx\pm 1$,
    but to identify how transitions occur we must separate the \emph{reaction direction} from the \emph{non-normal mode} that drives transient deviations~\cite{troude2024}. 

    \noindent\textbf{SVD approach.}
    Let $\PP_\pm = (\p_{+,\pm},\,\p_{-,\pm})$ be the eigenbasis matrix.  
    Its singular value decomposition reads $\PP_\pm=\U_\pm\Sig_\pm\V_\pm^\dag$,
    where $\U_\pm$ and $\V_\pm$ are unitary matrices,
    and $\Sig_\pm$ is a diagonal matrix composed of the singular value.
    We identify
    the column of $\U_\pm$ associated with the largest singular value as the reaction,  
    and the column associated with the smallest singular value as the non-normal mode.  

    To leading order in $C^{-1}$, $\PP_\pm$ is upper triangular, giving
    \begin{equation}
    \U_\pm = \frac{1}{\sqrt{2}}
    \begin{pmatrix}
    1 & 1 \\ 1 & -1
    \end{pmatrix}.
    \end{equation}
    Hence
    \(
    \rr = (1\,,\,1)/\sqrt{2}
    \),
    \(
    \nn =  (1\,,\,-1)/\sqrt{2}
    \).

    \noindent\textbf{Implications.}
    The reaction coordinate is therefore not aligned with the geometric axis $z_1=0$ connecting the equilibria.  
    Instead, we have to estimate if non-normal amplification pushes the system toward the separatrix,
    the curve separating the two basins of attraction.  

    Near the unstable equilibrium $\z_0$, the separatrix is tangent to the stable eigenvector $\p_{+,0}\approx (1\,,\,0)$,
    which is aligned along $z_1$, and the more the system gets close to the unstable equilibrium,
    the more the system is normal, and so non-normal amplification will not affect the stability of the system.
    In this limit, the $z_1$-axis remains stable, while the dynamics along $z_3$ is unstable, 
    so that transitions between the states $z_3 \approx +1$ and $z_3 \approx -1$ occur through 
    Brownian-like fluctuations. Thus, although the system is non-normal around stable equilibria,
    its linearized non-normality alone cannot guarantee accelerated switching, in this given model.

\subsection{Bistability and Non-Normal Acceleration}

    The original goal in \cite{zagkos2019} was to demonstrate the bistability of DNA methylation dynamics.  
    Their model explains the coexistence of hypo- and hypermethylated states but cannot account for experimentally observed \emph{fast} transitions (on the order of $\sim 10$ minutes) \cite{busto2020stochastic}.  
    In the original framework, transitions occur only near criticality,
    when a Jacobian eigenvalue crosses zero and the potential barrier vanishes,
    allowing noise to induce switching.
    \newline

    \noindent\textbf{Beyond criticality.}
    Our analysis combines three ingredients:
    \begin{enumerate}[(i)]
    \item the system is bistable, with two long-lived methylation states;
    \item transitions can occur at criticality, when a barrier disappears;
    \item even away from criticality, non-normality can amplify fluctuations, renormalizing the effective noise.
    \end{enumerate}
    To reconcile bistability with observed rapid switching, we propose the following modification 
    \begin{equation}    \label{eq:apx_z_dyn_nn}
    \begin{cases}
    \dot{z}_1 = -\omega_1 z_1 + \kappa^{-1}\beta (z_3-z_{3,+})(z_3-z_{3,-}) + \sqrt{2\delta}\,\eta_1, \\
    \dot{z}_3 = -\omega_3(z_3-z_{3,0})(z_3-z_{3,+})(z_3-z_{3,-}) + \kappa\beta z_1 + \sqrt{2\delta}\,\eta_3,
    \end{cases}
    \end{equation}
    with equilibria at $(0,z_{3,\pm})$ and unstable point $(0,z_{3,0})$, such that $z_{3,+}>z_{3,0}>z_{3,-}$.  
    This construction preserves the equilibria and their stability, but alters the flow structure so as to introduce genuine
    non-normal amplification, in line with Ref.~\cite{zagkos2019}. 
    This formulation reproduce the coexistence of long-term memory and fast stochastic
    transitions observed in methylation dynamics while preserving the equilibria and their stability.
    The dynamical system \eqref{eq:apx_z_dyn_nn} provides an illustrative case where we minimally modify an existing
    model to suggest how non-normal phase transitions could manifest in biology.
    More broadly, DNA methylation is paradigmatic because it simultaneously
    exhibits ``classical'' bistability (which secures epigenetic memory) and
    ``fast'' stochastic switching (which enables rapid adaptation). Our framework
    is unique in reconciling these two features. Furthermore, by linking the
    non-normality index $\kappa$ to the biochemical balance of DNMTs versus TET
    enzymes, the model acquires a direct mechanistic interpretation.
    In this way, we hope to attract the attention of the community to fully resolve the kinetics of DNA methylation
    from the non-normal dynamics perspective. 
    \newline

    \noindent\textbf{Interpretation.}
    \begin{itemize}
    \item Along $z_1$, the system is linearly stable, consistent with \cite{zagkos2019}.
    \item Along $z_3$, bistability arises from the cubic nonlinearity,
    with two stable equilibria and one unstable saddle.
    \item As $\omega_i \to 0^+$ or $z_{3,\pm}\to z_{3,0}$, the system approaches criticality.
    \item If $\kappa \ge \kappa_c = \omega_1/\beta$, the system enters a pseudo-critical regime:
    non-normality renormalizes the effective noise, enabling rapid switching even far from true criticality.
    \end{itemize}

\begin{figure}
        \centering
        \includegraphics[width=\textwidth]{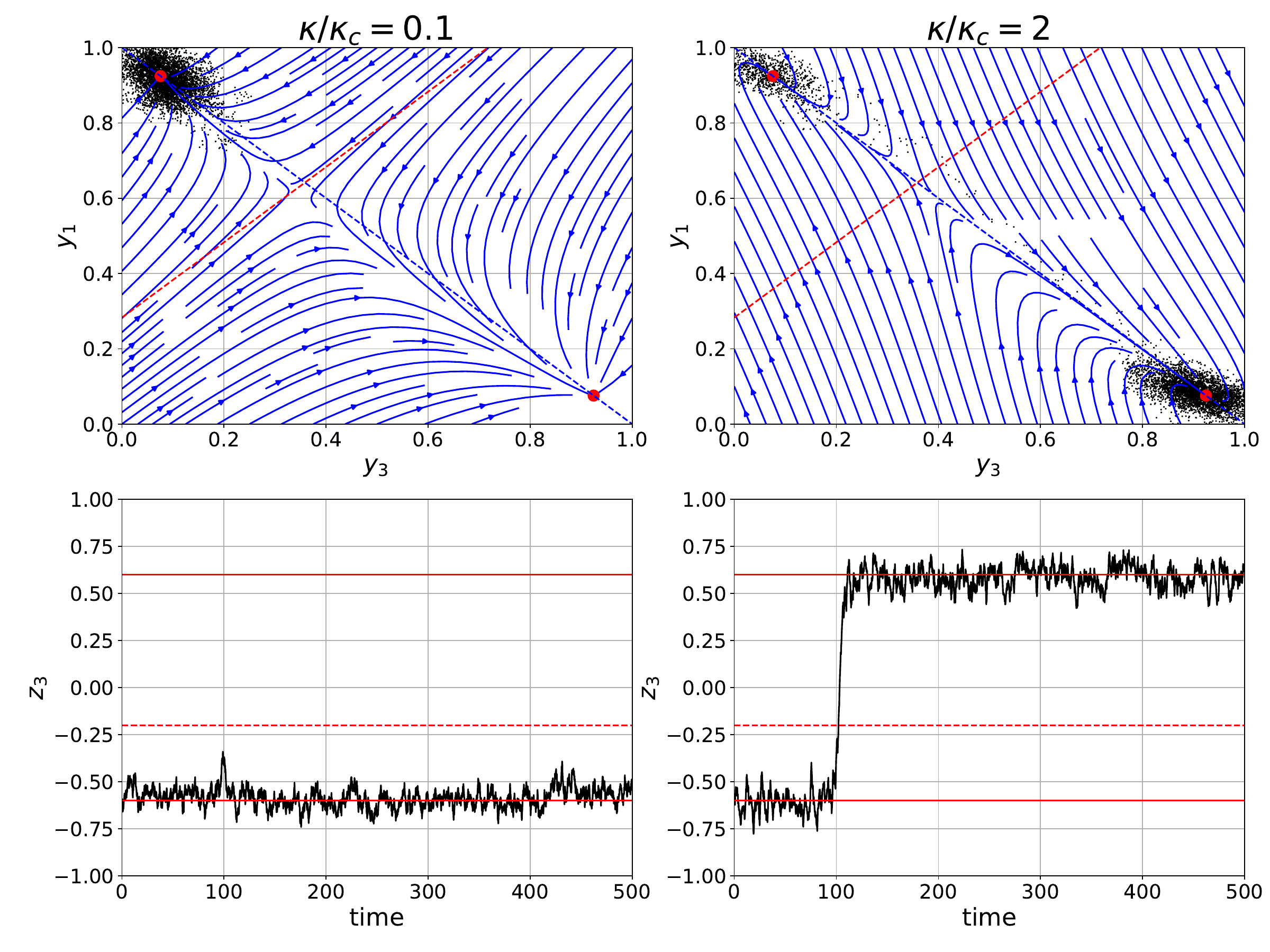}
        \caption{
            Simulation of a nonlinear two-dimensional system described in the $(z_1,z_3)$ by \eqref{eq:apx_z_dyn_nn}, 
            with parameters $z_{3,0}=-0.2$, $z_{3,+}=0.6$, $z_{3,-}=-0.6$ $\omega_1=\omega_3=1$, $\delta=0.001$, and $\kappa_c=10$. 
            The left panels correspond to $\kappa=\kappa_c/10$, while the right panels correspond to $\kappa=2\kappa_c$. \\
            Top panels: dynamics in phase space $(y_1,y_3)$ \eqref{eq:apx_z_to_y},
            where the horizontal axis denotes the ratio of methylated site ($y_3$) 
            and the vertical axis the ratio of unmethylated site ($y_1$). 
            Red dots mark the stable equilibria, 
            blue arrows indicate the force vector field, 
            and the dashed red line the axis $z_1$ and the blue dashed line the axis $z_3$,
            which crosses each other at the unstable equilibrium. \\
            Bottom panels: time series of the reaction variable ($z_3$). 
            Continuous red lines mark the stable equilibria at $z_{3,\pm}=\pm 0.6$, 
            and the dashed red line marks the unstable equilibrium at $z_{3,0}=-0.2$. \\
            All simulations are performed over a time horizon $T=500$ 
            with integration step $\Delta t=0.1$, 
            corresponding to $N=5000$ time steps.
        }
        \label{fig:phase_dnmt}
    \end{figure}

    \noindent\textbf{Numerical Analysis.}
    In Figure~\ref{fig:phase_dnmt}, we show two simulations of the ratio of methylated and unmethylated sites, 
    i.e. the $(y_1,y_2)$ space defined in \eqref{eq:apx_z_to_y}, obtained by simulating \eqref{eq:apx_z_dyn_nn} 
    with $\kappa = \kappa_c/10$ and $\kappa = 2\kappa_c$. 
    To introduce asymmetry, we choose parameters such that 
    $|z_{3,+}-z_{3,0}| > |z_{3,-}-z_{3,0}|$. 
    When $\kappa < \kappa_c$, the system remains stable around both equilibria. 
    However, as $\kappa$ approaches $\kappa_c$, the system exits the stable equilibrium $z_{3,-}$ 
    and becomes trapped around the second stable equilibrium $z_{3,+}$. 
    This occurs because the mean-reversion rate of the dynamics of $z_3$ near $z_{3,+}$ is stronger than near $z_{3,-}$, 
    as the potential barrier between $z_{3,+}$ and $z_{3,0}$ is higher than the one between $z_{3,-}$ and $z_{3,0}$.

\begin{figure}
        \centering
        \includegraphics[width=\textwidth]{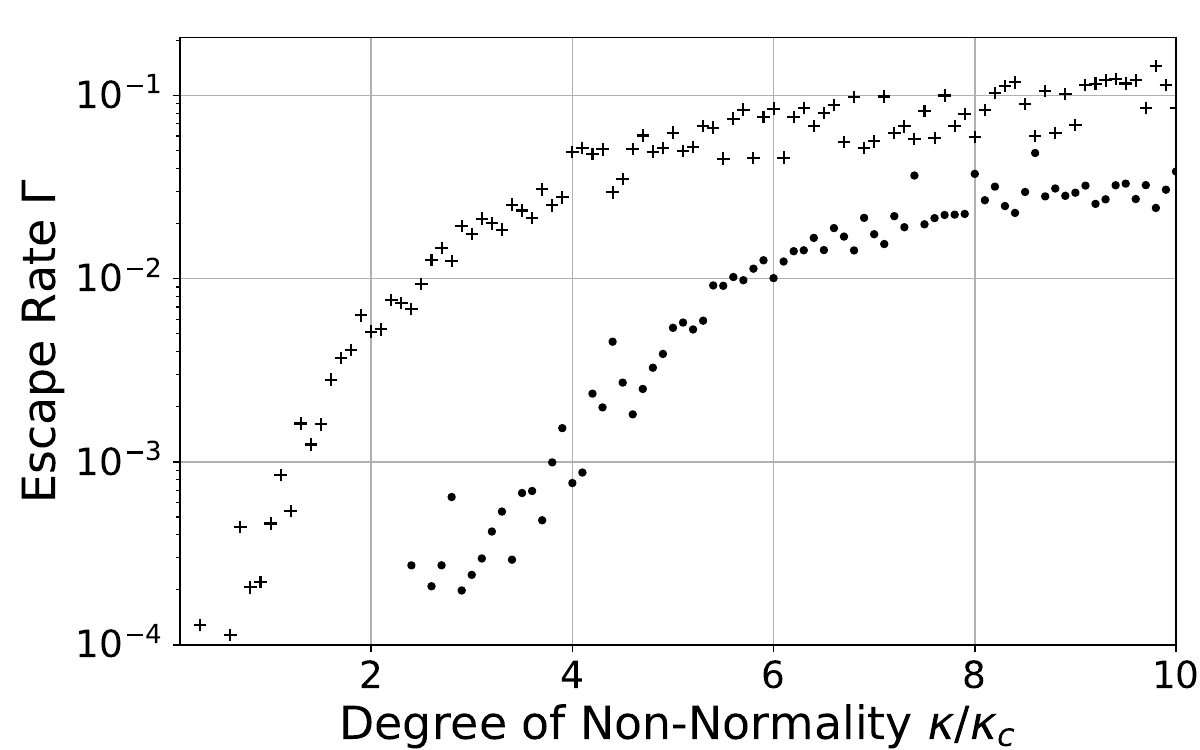}
        \caption{
            Escape rate of the reaction variable $z_3$ as a function of $\kappa/\kappa_c$.  
Plain dots (``$\cdot$'') denote transitions from $z_3>z_{3,0}$ to $z_3<z_{3,0}$,  
while crosses (``$+$'') denote the reverse transitions.  
The dynamics follow \eqref{eq:apx_z_dyn_nn} with parameters  
$z_{3,0}=-0.2$, $z_{3,+}=0.6$, $z_{3,-}=-0.6$, $\omega_1=\omega_3=1$, $\delta=0.001$, and $\kappa_c=10$.  
Simulations are performed for $\kappa$ ranging from $\kappa_c/10$ to $10\kappa_c$,  
each run covering a total time $T=10^{5}$ with integration step $\Delta t=0.1$,  
corresponding to $N=10^{5}$ sampled points.  
The escape rate is computed as $\Gamma = 1/\langle\tau\rangle$,  
where $\langle\tau\rangle$ is the mean first-passage time across the unstable saddle $z_{3,0}$.       }
        \label{fig:rate_dnmt}
    \end{figure}

    This asymmetry between the two transition rates is made explicit in Figure~\ref{fig:rate_dnmt}, 
    where we plot the measured transition rates from each equilibrium as a function of $\kappa/\kappa_c$. 
    For instance, when $\kappa = 4\kappa_c$, the transition rate from $z = z_{3,-}$ to $z_{3,+}$ 
    in the considered time unit is $\Gamma_{z_{3,-}\to\z_{3,-}} \approx 5 \times 10^{-2}$, 
    whereas the reverse rate is only $\Gamma_{z_{3,+}\to z_{3,-}} \approx 10^{-5}$. 
    Thus, non-normality can explain the rapid transitions between states, 
    but such transitions are not necessarily reversible due to the asymmetry of the potential landscape.

    \noindent\textbf{Conclusion.}
    Non-normality thus reconciles bistability with rapid dynamics:  
    DNA methylation can be both stable (supporting epigenetic memory) and fast-adapting (enabling minute-scale responses) through transient amplification of stochastic fluctuations.

\subsection{Conclusion}

    We have shown how an existing nonlinear model of CpG dyads~\cite{zagkos2019} 
    can be extended with explicit stochasticity and analyzed through the lens of non-normality.  
    This approach reveals how a purely deterministic bistable model,
    once augmented by a non-normal control parameter $\kappa$,  
    can amplify thermal fluctuations and thereby accelerate transitions between unmethylated and methylated states 
    without altering the system's spectral stability.  
    In this way, non-normality provides a mechanistic explanation for the rapid DNA methylation dynamics observed experimentally~\cite{busto2020stochastic}, 
    while preserving the underlying bistability that supports epigenetic memory.

    \noindent\textbf{Biological interpretation of $\kappa$.}
    The parameter $\kappa$ quantifies the strength of non-normal amplification in our reduced model.  
    Biologically, it integrates the balance between DNA methyltransferases (DNMTs) and TET demethylases.  
    DNMT3a and DNMT3b establish new methylation marks, DNMT1 maintains them during replication, 
    while TET enzymes actively remove them via iterative oxidation of 5-methylcytosine.  
    Thus, elevated DNMT activity or reduced TET activity can correspond to high $\kappa$, 
    whereas the converse produces low $\kappa$.

    \noindent\textbf{Low-$\kappa$ regimes.}
    If $1 < \kappa < \kappa_c$, fluctuations are not strongly amplified,
    and fare from the criticality the system stays asymptotically stable.
    The only way for the system to transit between equilibrium, is to spectral criticality.

    \noindent\textbf{High-$\kappa$ regimes.}
    When $\kappa \gtrsim \kappa_c$, non-normality strongly renormalizes effective noise, enabling rapid switching.  
    This regime can arise through:
    \begin{itemize}
    \item DNMT overexpression or TET downregulation: observed in several cancers~\cite{Klutstein2016,McGovern2012}, 
    leading to accelerated conversion of hemimethylated sites into fully methylated ones.
    \item Efficient maintenance methylation: when DNMT1 rapidly restores methylation after replication and demethylation processes are weak, 
    as observed in certain adult tissues and tumor cell lines~\cite{McGovern2012}.
    \item TET hyperactivity or reduced DNMT expression: producing a net bias toward demethylation, 
    as seen in promoters of constitutively hypomethylated genes~\cite{Li2010,Hunter2012}.
    \item Passive demethylation: inefficient DNMT1 activity, particularly during aging, 
    leads to progressive loss of methylation across divisions~\cite{Li2010}.
    \end{itemize}
    The difference between the regime can be quantify by the asymmetry in the potential \eqref{eq:apx_z_dyn_nn}.

    \noindent\textbf{Closing.}
    In summary, DNA methylation provides a compelling application of our extended non-variational Kramers framework.  
    The degree of non-normality $\kappa$ can, in principle, be inferred from empirical measurements of methylation state transitions.  
    Evidence from both normal and pathological contexts suggests that methylation dynamics frequently operate in the high-$\kappa$ regime, 
    where transient amplification drives rapid state switching.  
    By extending the model in \cite{zagkos2019} to include Gaussian noise and non-normal amplification, 
    we reconcile bistability with fast epigenetic responses, 
    providing a theoretical basis for the sudden methylation transitions observed in vivo~\cite{busto2020stochastic}.

\section{Overdamped Kramer Escape Rate}
\label{apx:kramer}

    In this section, we introduce the mathematical framework used to derive the escape rate in the overdamped limit.
    We are borrowing from the derivation made by H.A. Kramer (1940) \cite{kramers1940} of the escape rate of a particle in a one dimensional potential well in the overdamped limit.
  
    We consider a system described by $x$, evolving in a potential $U(x)$
    with a minimum at $x_i$ and a potential barrier at $x_f$.
    Therefore, in the overdamped limit, we can write a one-dimensional Langevin equation as
    \begin{equation}
        \dot{x} = - U'(x) + \sqrt{2\delta}\eta(t)
        .
    \end{equation}
    For this problem, we know that the probability density function $P(x,t)$ satisfies the Fokker-Planck equation
    \begin{subequations}
        \begin{align}
            &\partial_t P = \partial_x\left[U'(x)P\right] + \delta\partial_x^2 P = -\partial_x J
            , \\
            \text{where}\;
            &J(x,t) = -U'(x)P - \delta\partial_x P 
        \end{align}
    \end{subequations}
    is the probability current.
    If the probability is constant and the current is equal to zero ($J(x,t)=0$),
    the solution of the Fokker-Planck equation is given by the Boltzmann distribution
    i.e. $P(x)\sim e^{-U(x)/\delta}$. 

    To obtain the escape rate of the particle from the potential well,
    we search for an almost stationary solution of the Fokker-Planck equation
    i.e. $\partial_t P \approx 0$, which allows us to assume that the probability current is almost constant and uniform
    i.e. $J(x,t)=J$.
    This leads to
    \begin{equation}
        J
        = -U'(x)P - \delta\partial_x P
        = -\delta e^{-\frac{U(x)}{\delta}}\partial_x\left[e^{\frac{U(x)}{\delta}} P\right]
    \end{equation}
    \begin{equation}
        \Rightarrow\quad
        \partial_x\left[e^{\frac{U(x)}{\delta}} P\right] = \frac{J}{\delta}e^{\frac{U(x)}{\delta}}
        .
    \end{equation}
    Integrating the last equation from the bottom of the potential well at $x_i$ to a point $x'$,
    even beyond the potential barrier at $x_f$, and assuming
    that the probability density is almost zero at $x'$,
    the probability current is obtained from
    \begin{subequations}
        \begin{align}
            \frac{J}{\delta}\int^{x'} _{x_i} e^{\frac{U(x)}{\delta}}\, \mathrm{d}x
            &= e^{\frac{U(x')}{\delta}}P[x=x'] - e^{\frac{U(x_i)}{\delta}}P[x=x_i] \\
            &\approx - e^{\frac{U(x_i)}{\delta}}P[x=x_i]
            \quad\text{since } P[x=x']\approx 0 ,
        \end{align}
    \end{subequations}
    \begin{equation}    \label{eq:current}
        \Rightarrow\quad
        J \approx \delta\frac{P\left[x=x_i\right]e^{\frac{U(x_i)}{\delta}}}{\int_{x_i}^{x'} e^{\frac{U(x)}{\delta}}dx}
        .
    \end{equation}

    The escape rate $\Gamma$ is given by the probability current per unit of time,
    conditional to having the particle in the well.
    Denoting the probability that the particle is in the well as $p_0$,
    the probability current is $J = p_0 \Gamma$.
    Under the hypothesis that the barrier is high enough,
    the probability that the particle is in the well can be approximated by
    \begin{align}
        p_0   &= \int^{x_i+ \delta}_{x_i-\delta}P(x)\, \mathrm{d}x \\
            &\approx P[x=x_i]\int^{x_i+\delta}_{x_i-\delta}e^{-\frac{1}{\delta}(U(x) - U(x_i))}\, \mathrm{d}x  \\
            &\approx P[x=x_i]\int^{x_i+\delta}_{x_i-\delta}e^{-\frac{1}{2\delta}U''(x_i)x^2}\, \mathrm{d}x  \\
            &\approx P[x=x_i]\int^{+\infty}_{-\infty}e^{-\frac{1}{2\delta}U''(x_i)(x-x_i)^2}\, \mathrm{d}x  \\
        \Rightarrow\; p_0
        &\approx P[x=x_i]\sqrt{\frac{2\pi\delta}{U''(x_i)}}
            \quad\text{.}
    \end{align}
    On the other hand, the integral in the denominator of the probability current \eqref{eq:current} 
    can be approximated by
    \begin{align}
        \int^{x'}_{x_i}e^{\frac{1}{\delta}U(x)}\, \mathrm{d}x
        &\approx e^{\frac{1}{\delta}U(x_f)}\int^{x'}_{x_i}e^{\frac{1}{2\delta}U''(x^*)(x-x_f)^2}\, \mathrm{d}x \\
        &\approx e^{\frac{1}{\delta}U(x_f)}\int^{+\infty}_{-\infty}e^{-\frac{1}{2\delta}|U''(x_f)|(x-x_f)^2}\, \mathrm{d}x \\
        &\approx \sqrt{\frac{2\pi\delta}{|U''(x^*)|}}e^{\frac{1}{2\delta}U(x_f)}
        \quad\text{.}
    \end{align}
    We thus obtain the escape rate as
    \begin{equation}
        \Gamma = \frac{1}{2\pi}\sqrt{U''(x_i)|U''(x_f)|} ~~e^{-\frac{\Delta E}{2\delta}}
        ,
    \end{equation}
    where $\Delta E = U(x_f) - U(x_i)$ is the height of the potential barrier.

\end{document}